\documentclass[twocolumn]{aastex701}

\renewcommand{\textbf}{}

\usepackage{amsmath}
\numberwithin{equation}{section}
\usepackage{tabularx}   
\usepackage{array}      
\usepackage{makecell}   
\usepackage{multirow}   
\usepackage{ragged2e}   
\usepackage{array}
\usepackage{ragged2e}
\usepackage{makecell}
\usepackage{here}
\usepackage{cancel}

\begin{document}

\title{Radial evolution of Alfvén wave Parametric Decay Instability in the near-Sun solar wind:\\ Effects of Temperature Anisotropy}

\author[orcid=0009-0001-1813-8302]{Hayato Saguchi}
\affiliation{Department of Geophysics, Graduate School of Science, Tohoku University, 6-3 Aramaki-Aza-Aoba, Aoba, Sendai, Miyagi 980-8578, Japan}
\email[show]{saguchi.hayato.s4@dc.tohoku.ac.jp}  

\author[orcid=0000-0002-8787-5170]{Yohei Kawazura}
\affiliation{School of Data Science and Management, Utsunomiya University, 350 Minemachi, Utsunomiya, Tochigi 321-8505, Japan}
\affiliation{Department of Geophysics, Graduate School of Science, Tohoku University 6-3 Aoba, Aramaki, Sendai 980-8578, Japan}
\email[hide]{kawazura@a.utsunomiya-u.ac.jp}  

\author[orcid=0000-0002-7136-8190]{Munehito Shoda}
\affiliation{Department of Earth and Planetary Science, School of Science, The University of Tokyo, 7-3-1 Hongo, Bunkyo-ku, Tokyo, 113-0033, Japan}
\email[hide]{shoda.m.astroph@gmail.com}

\author[orcid=0000-0002-4318-0633]{Yuto Katoh}
\affiliation{Department of Geophysics, Graduate School of Science, Tohoku University, 6-3 Aramaki-Aza-Aoba, Aoba, Sendai, Miyagi 980-8578, Japan}
\email[hide]{yuto.katoh@tohoku.ac.jp}  

\begin{abstract}

Parametric decay instability (PDI) of Alfvén wave is thought to play an important role in the dissipation of the large-amplitude Alfvén waves and in the heating of magnetized plasmas. Temperature anisotropy is frequently observed by spacecraft, including Parker Solar Probe (PSP), in the near-Sun solar wind, yet its impact on PDI in the near-Sun solar wind has been understudied. We calculate the maximum growth rates of PDI, $\gamma_{\max}/\omega_{0}$, where $\omega_0$ is the frequency of the parent wave, by solving the linear dispersion relation of Chew-Goldberger-Low (CGL) equations under several expanding background models. To assess the effect of temperature anisotropy, the growth rate is compared with that derived from ideal magnetohydrodynamics (MHD). From $R_0$ ($ = 1.02R_\odot$) to $30R_0$, we consider three expansion cases: (i) spherically symmetric adiabatic expansion with constant wind speed, (ii) Multi-source observation- and model-constrained expansion, and (iii) a PSP-constrained profile of $(\beta_{\parallel},\xi)$, where $\beta_\parallel=8\pi p_{\parallel0}/B_0^2$ is the parallel plasma beta and $\xi=T_{\perp0} / T_{\parallel0}$ is the temperature anisotropy, that includes Parker-spiral effects. We find that temperature anisotropy increases $\gamma_{\max}/\omega_{0}$ for $\beta\lesssim 0.1$ in the near-Sun solar wind: in the case of (iii), temperature anisotropy with $T_{\perp0} > T_{\parallel0}$ increases $\gamma_{\max}/\omega_{0}$ by factors of $\sim 1.5$ over $R\simeq 1$--$10\,R_0$, whereas temperature anisotropy with $T_{\parallel0}>T_{\perp0}$ decreases $\gamma_{\max}/\omega_{0}$ at larger $R$. 
Our results suggest that the temperature anisotropy plays an important role in the onset of PDI even in low-$\beta$ regimes, such as the near-Sun solar wind.

\end{abstract}



\section{Introduction} 
One of the major unresolved problems in solar physics is how the solar corona is heated and how the solar wind is accelerated. It is widely believed that the energy source of these phenomena comes from energy of magneto-convection on the photosphere, and Alfvén waves carry the energy to heat the solar corona and drive the solar wind. In fact, some studies suggest that Alfvén waves observed in the chromosphere and low corona have enough energy to drive the solar wind \citep{DePontieuEtAl2007, McIntoshEtAl2011}. The mechanism of how Alfv\'en waves propagate and dissipate is, however, still under discussion. 

Reflection-driven Alfv\'en wave turbulence has been extensively studied as a candidate process for coronal heating and the solar wind acceleration by Alfv\'en waves \citep[e.g.,][]{MatthaeusEtAl1999, DmitrukEtAl2002, CranmerVanBallegooijenEdgar2007, VerdiniVelli2007, PerezChandran2013, vanderHolstEtAl2014, vanBallegooijenAsgariTarghi2016, vanBallegooijenAsgariTarghi2017, UsmanovEtAl2018, ChandranEtAl2025b, MeyrandSquireMalletChandran2025}. Due to the gradients of the background Alfvén speed in the solar corona and the solar wind, Alfv\'en waves are partially reflected \citep{HeinemannOlbert1980, Velli1993}. The nonlinear interaction between sunward and anti-sunward propagating Alfvén waves occurs, leading to the onset of Alfv\'en wave turbulence, which drives energy to small scales, ultimately heating the plasma. \citep{HowesNielson2013}. Previous studies suggest that Alfv\'en wave turbulence model based on reduced MHD model may be insufficient to provide enough energy to sustain the solar wind \citep[e.g.,][]{PerezChandran2013}, whereas small-scale gradients due to density fluctuations can increase the heating rate \citep{vanBallegooijenAsgariTarghi2016, vanBallegooijenAsgariTarghi2017}. Indeed, reduced MHD model that incorporates observationally constrained density fluctuations showed sufficient Alfvén wave reflection and turbulence to heat coronal holes and drive the fast solar wind \citep{AsgariTarghiEtAl2021}. From this perspective, several studies have included compressibility into Alfvén wave turbulence model and have succeeded in reproducing coronal heating and the fast solar wind self-consistently \citep{ShodaYokoyama2018, ShodaYokoyamaSuzuki2018a, ShodaYokoyamaSuzuki2018b, ShodaEtAl2019, Matsumoto2021}.

As a physical process for generating the density fluctuation, enhancing reflection rate and promoting Alfvénic turbulence \citep[e.g.,][]{SuzukiInutsuka2006, ShodaEtAl2019, Matsumoto2021}, parametric decay instability (PDI) of Alfvén wave has attracted attention. In PDI, a large-amplitude forward propagating Alfv\'en wave (referred to as mother wave, parent wave, or pump wave) decays into daughter waves: a backward propagating Alfv\'en wave and a forward propagating slow-magnetoacoustic wave \citep[][]{GaleevOraevskii1963, SagdeevGaleev1969}. \cite{Goldstein1978} and \cite
{Derby1978} derived the dispersion relation of PDI from ideal MHD equations. The solar corona and the near-Sun solar wind are a favorable environment for the development of PDI because the instability preferentially occurs in low-$\beta$ plasmas.

PDI has been studied extensively over the past several decades including dispersive effects \citep[e.g.,][]{WongGoldstein1986, LongtinSonnerup1986, Hollweg1994}, kinetic effects \citep[e.g.,][]{Inhester1990, Araneda1998, NariyukiHada2006, NariyukiHada2007}, two-fluid \citep[e.g.,][]{VinasGoldstein1991a, VinasGoldstein1991b}, expansion effects \citep[e.g.,][]{MatteiniLandiVelliHellinger2010, TeneraniVelli2013}, the suppressing effects of turbulence on PDI \citep[e.g.,][]{Shi2017, Fu2018}, temperature anisotropy \citep[e.g.,][]{Hamabata1993,Tenerani2017}, relativistic effects \citep{IshizakiIoka2024}, and the parametric decay of kinetic Alfv\'en waves \citep[e.g.,][]{HasegawaChen1976, LiFuDorfman2024}. PDI has also been investigated in a variety of settings and with diverse numerical approaches, including studies of nonlinear evolution \citep[e.g.,][]{TerasawaEtAl1986, HoshinoGoldstein1989}, non-monochromatic waves \citep[e.g.,][]{UmekiTerasawa1992, MalaraVelli1996}, multi-dimensional numerical simulations \citep[e.g.,][]{DelZanna2001, DelZanna2015}, fully kinetic particle-in-cell (PIC) simulations \citep[e.g.,][]{NariyukiMatsukiyoHada2008, GonzalezInnocentiTenerani2023}, and Alfv\'enic wave packets \citep[e.g.,][]{TeneraniEtAl2020, MarriottTenerani2024, MarriottTenerani2024b, Komissarov2025}. In addition, several studies have examined secondary effects triggered by PDI, such as inverse cascade \citep[e.g.,][]{Reville2018, Chandran2018}, anisotropic heating via stochastic heating \citep[][]{Comisel2019}, and PDI-driven turbulence \citep[e.g.,][]{ShodaYokoyama2018, ShodaEtAl2019}. Several studies have attempted to identify observational evidence of PDI. The signatures of PDI were reported in the solar transition region \citep{Hahn2022}, and several other works have presented suggestive results \citep{Bowen2018, KasperEtAl2021, HahnEtAl2025, GonzalezEtAl2026}. Although \cite{LiDorfmanFu2025} proposed a novel and experimentally viable scheme to quantify the growth of PDI recently, observational examples of PDI remain relatively scarce \citep[][]{Zank2022, Zhao2022}.

Temperature anisotropy is frequently observed by Parker Solar Probe (PSP) in the near-Sun solar wind (e.g., \citealt{Verniero2020, Woodham2021, Huang2020, Huang2023a, Huang2023b, Huang2025, Short2024, Laker2024, Yogesh2025, Coello-Guzman2026}). In some intervals, the observed anisotropy becomes high, with $T_{\perp}/T_{\parallel} \gtrsim 10$ around the ion cyclotron instability threshold \citep[e.g., ][]{Huang2023b}, where $\perp$ and $\parallel$ denote the perpendicular and parallel components to the magnetic field. Theoretical modeling of species-dependent temperatures and temperature anisotropies in the solar wind remains an active and important topic \citep[e.g.,][]{OfmanVinasGary2001, OfmanGaryVinas2002, Meng2015, ChandranEtAl2025a, Yoon2026}. In particular, several mechanisms have been proposed to account for them, such as the helicity barrier \citep[e.g.,][]{Meyrand2021, Squire2022, McIntyre2025, Panchal2025} and the stochastic heating, both of which have attracted considerable attention as an efficient channel for perpendicular heating in the solar wind \citep[e.g.,][]{Chandran2010, Chandran2013, Cerri2021, Bowen2025, Johnston2025, Mallet2026}. In global solar-wind modeling, Alfvén Wave Solar Model (AWSoM; \citealt{vanderHolstEtAl2014, Meng2015, vanderHolst2022}) has achieved notable success in predicting temperature anisotropies via stochastic heating.

Regarding the relationship between temperature anisotropy and PDI, however, theoretical studies that explicitly quantify its effects on PDI are scarce. The pioneering analyses of \citet{Hamabata1993} and \citet{Tenerani2017} employ linear theory based on the Chew-Goldberger-Low (CGL) equations \citep{ChewGoldbergerLow1956} and show that a background temperature anisotropy enhances the growth rate when $T_{\perp} > T_{\parallel}$ and reduces it when $T_{\perp} < T_{\parallel}$. Nevertheless, these works do not address the regime most relevant to the near-Sun solar wind—namely, low-$\beta$ conditions in which PSP observes large temperature anisotropies—leaving its impact on the growth rate largely unexplored. Furthermore, the heliocentric locations where PDI is expected to occur have been investigated by computing the radial evolution of the maximum growth rate in expanding isotropic solar-wind models \cite[e.g., ][]{UmekiTerasawa1992, TeneraniVelli2013, DelZanna2015, Reville2018, ShodaYokoyamaSuzuki2018b, ShodaEtAl2019}. However, the radial evolution of the maximum growth rate based on PSP observations remains poorly explored, and, to date, no study has incorporated temperature anisotropy within such a framework. As a result, our understanding of how the pronounced near-Sun anisotropy influences the onset and radial evolution of PDI remains unclear.


In this study, following \cite{Tenerani2017}, we analyze the maximum growth rate of PDI in the near-Sun solar wind, incorporating temperature anisotropy. Specifically, we address the following questions.
\begin{enumerate}
    \item How much does the temperature anisotropy enhance the maximum growth rate of PDI in the low-$\beta$ region?
    \item In the adiabatic expansion, does the temperature anisotropy affect the radial evolution of the maximum growth rate of PDI?
    \item In the expansion cases based on the observed data of PSP, does the temperature anisotropy affect the radial evolution of the maximum growth rate of PDI? 
\end{enumerate}
Additionally, we discuss the effect of the perturbed temperature anisotropy on the maximum growth rate of PDI.

\section{method}\label{sec:style}


\subsection{Governing equations and dispersion relations} \label{sec:style}

We use two dispersion relations of the parametric instability that are derived respectively from their basic equations, the isotropic one (ideal MHD) and the anisotropic one 
\citep[CGL equations,][]{ChewGoldbergerLow1956}
to compare two results and evaluate the effect of temperature anisotropy. Dispersion relations can be derived by linearizing the governing equations about a finite-amplitude mother wave and retaining small perturbations corresponding to the daughter waves and the background plasma \citep{Goldstein1978, Derby1978}. The governing equations in the isotropic case (ideal MHD) are as follows:
\begin{equation}
\frac{\partial \rho}{\partial t} + \nabla\cdot(\rho \mathbf{u}) = 0, 
\label{M1}
\end{equation}

\begin{equation}
\rho\left(\frac{\partial}{\partial t} + \mathbf{u}\cdot\nabla\right)\mathbf{u}
= -\nabla p + \frac{(\nabla\times\mathbf{B})\times\mathbf{B}}{4\pi},
\label{M2}
\end{equation}

\begin{equation}
\frac{\partial \mathbf{B}}{\partial t}
= \nabla\times(\mathbf{u}\times\mathbf{B}),
\label{M3}
\end{equation}

\begin{equation}
\frac{dp}{d\rho} = c_s^{2}.
\label{M4}
\end{equation}

\begin{equation}
    p \rho^{-n} = \rm{const},
\label{M5}
\end{equation}
where the parameters in a set of equations are as follows:
\begin{itemize}
    \item $\rho$ denotes the mass density.
    \item $\mathbf{u}$ denotes the vector of the velocity.
    \item $p$ denotes the pressure.
    \item $\mathbf{B}$ denotes the vector of the magnetic field.
    \item $c_s$ denotes the sound speed.
    \item $n$ denotes the polytropic index. In this study, we assume adiabaticity, so we set $n$ as $5/3$.
\end{itemize}
After linearization, the dispersion relation is as follows \citep{Goldstein1978, Derby1978}: 
\begin{equation}
\begin{split}
(\hat{\omega}-\hat{k})\left(\hat{\omega}^{2}-\frac{5}{6}\beta\hat{k}^{2}\right)
\left[(\hat{\omega}+\hat{k})^{2}-4\right] \\
= \hat{B}_{\perp}^{2}\hat{k}^{2}\left(\hat{\omega}^{3}+\hat{k}\hat{\omega}^{2}-3\hat{\omega}+\hat{k}\right).
\end{split}
\label{M6}
\end{equation}
where the parameters in the dispersion relation are as follows:
\begin{itemize}
  \item $\hat{\omega} \equiv \omega/\omega_{0} + i\,\gamma/\omega_{0}$, \\
  the (dimensionless) daughter-wave complex frequency, normalized by the mother-wave frequency $\omega_{0}$.

  \item $\hat{k} \equiv k/k_{0}$, \\
  the (dimensionless) daughter-wave wavenumber, normalized by the mother-wave wavenumber $k_{0}$.

  \item $\beta \equiv 8\pi p_0/B_0^2$, \\
  the plasma beta, defined as the ratio of the gas pressure to the magnetic pressure.

  \item $v_{A} \equiv B_{0}/\sqrt{4\pi\rho_{0}}$, \\
  the Alfv\'en speed defined in terms of the background magnetic-field strength $B_{0}$ and mass density $\rho_{0}$.

  \item $\hat{B}_{\perp} \equiv B_{\perp}/B_{0}$, \\
  the mother-wave transverse magnetic-field amplitude $B_{\perp}$ normalized by the background field $B_{0}$.
\end{itemize}





In the anisotropic case, we adopt the Chew–Goldberger–Low (CGL) equations, which describe temperature anisotropy in a weakly collisional plasma under double-adiabatic conditions \citep{ChewGoldbergerLow1956}. CGL equations are as follows:
\begin{equation}
\frac{\partial \rho}{\partial t} + \nabla\cdot(\rho \mathbf{u}) = 0,
\label{M7}
\end{equation}
\begin{equation}
\begin{split}
\rho\left(\frac{\partial \mathbf{u}}{\partial t} + \mathbf{u}\cdot\nabla \mathbf{u}\right)
&= -\nabla\left(p_\perp + \frac{B^2}{8\pi}\right) \\
&\quad + \frac{1}{4\pi}\,\mathbf{B}\cdot\nabla\mathbf{B}
      + \nabla\cdot\!\left(\hat{\mathbf{b}}\hat{\mathbf{b}}\,\Delta p\right),
\end{split}
\label{M8}
\end{equation}
\begin{equation}
\frac{\partial \mathbf{B}}{\partial t} = \nabla\times(\mathbf{u}\times\mathbf{B}),
\label{M9}
\end{equation}
\begin{equation}
\left(\frac{\partial}{\partial t} + \mathbf{u}\cdot\nabla\right)
\left(\frac{p_\parallel B^2}{\rho^3}\right) = 0,
\label{M10}
\end{equation}
\begin{equation}
\left(\frac{\partial}{\partial t} + \mathbf{u}\cdot\nabla\right)
\left(\frac{p_\perp}{\rho B}\right) = 0,
\label{M11}
\end{equation}
where the new parameters in a set of equations are as follows:
\begin{itemize}
    \item $p_\perp$ denotes the pressure perpendicular to the magnetic field.
    \item $p_\parallel$ denotes the pressure parallel to the magnetic field.
    \item $\mathbf{\hat{b}}=\mathbf{B}/B$ denotes the unit vector along the magnetic field.
    \item $\Delta p = p_\perp - p_\parallel$ denotes the pressure anisotropy, the difference between $p_\perp$ and $p_\parallel$.
    
\end{itemize}
After linearization, the dispersion relation derived from CGL equations \citep{Tenerani2017} is as follows:
\begin{widetext}
\begin{equation}
\begin{split}
\left[\hat{\omega}^{2}-\tilde{\beta}\hat{k}^{2}\left(1+\frac{\hat{B}_{\perp}^{2}\xi}{3}\right)\right]
\left\{(\hat{\omega}-\hat{k})\left[(\hat{\omega}+\hat{k})^{2}-4\right]
+\frac{\tilde{\beta}\hat{B}_{\perp}^{2}(\xi-4)}{3\left(1+\hat{B}_{\perp}^{2}\right)}
\left[(\hat{k}^{2}+1)\hat{\omega}+\hat{k}(\hat{k}^{2}-3)\right]\right\} \\
= \hat{B}_{\perp}^{2}\hat{k}^{2}\left[1-\frac{\tilde{\beta}\left(3-\xi-\hat{B}_{\perp}^{2}\right)}{3\left(1+\hat{B}_{\perp}^{2}\right)}\right]
\left\{\hat{\omega}^{3}+\hat{\omega}^{2}\hat{k}-3\hat{\omega}+\hat{k}
-\frac{\tilde{\beta}(3-\xi)}{3}
\left[(\hat{k}^{2}+1)\hat{\omega}+\hat{k}(\hat{k}^{2}-3)\right]\right\},
\end{split}
\label{M12}
\end{equation}
\end{widetext}
where the plasma parameters in the dispersion relation are as follows:
\begin{itemize}
    \item $\hat{\omega}, \hat{k}$ and $\hat{B}_{\perp}$ are the same as those of the ideal-MHD case.
    \item $\beta_{\parallel}=8\pi p_{\parallel0}/B_{0}^2$, \\ the parallel plasma beta.
    \item $\xi=T_{0\perp}/T_{\parallel0}$, \\ the background temperature anisotropy.
    \item $ \tilde{\beta}=3\beta_{\parallel}/[2(1+\hat{B}_{\perp}^2+\beta_{\parallel}(\xi-1)/2)]$.
\end{itemize}
Note that we do not solve the full set of governing equations numerically, but solve two dispersion relations by prescribing the background model beforehand in this study. 

\begin{deluxetable*}{cccc}
\tablewidth{0pt}
\tabletypesize{\normalsize}
\tablecaption{\normalsize Background solar wind models \label{tab:description}}
\tablehead{
\colhead{} & \colhead{Evolution of $B_0$ and $\rho_0$} & \colhead{Evolution of $T_{\perp0}$ and $T_{\parallel0}$} & \colhead{Evolution of $\hat{B}_{\perp}^2$}
}
\startdata
\makecell[c]{Case 1\\(Spherically symmetric adiabatic expansion \\ with constant wind speed)} & \makecell[c]{$B_0 \propto R^{-2}$ \\ $\rho_0 \propto R^{-2}$} & \makecell[c]{Anisotropic: \\ $\frac{d}{dt}(p_{\parallel0}B_{0}^2/\rho_{0}^3)=0$ \& $\frac{d}{dt}(p_{\perp0}/\rho_{0} B_{0})=0$ \\ $\rightarrow \ T_{\parallel0}=const, \ T_{\perp0} \propto R^{-2}$ \\ Isotropic: \\ $T_{0} \rho_{0}^{-2/3}=const \rightarrow \ T_{0} \propto R^{-4/3}$} & \makecell[c]{$\hat{B}_{\perp}^2 \propto R^1$} \\ \tableline
\makecell[c]{Case 2 \\  (Multi-source observation- and \\ model-constrained expansion)} & \makecell[c]{Profiles derived from \\ the analysis of Metis \\ and PSP observations\\ (\cite{Telloni2021})} & \makecell[c]{Anisotropic: \\ The result of AWSoM \\  (\cite{Meng2015}) \\ Isotropic: \\ $T_0=(2T_{\perp0}+T_{\parallel0})/3$ \\ with the anisotropic profile} & \makecell[c]{$\hat{B}_{\perp}^2 \propto R^{0.96}$ \\ (\cite{Huang2023c})} \\ \tableline
\makecell[c]{Case 3 \\ (PSP-constrained expansion)} & \makecell[c]{Profiles derived from \\ PSP observations\\ (\cite{Short2024})} & \makecell[c]{Anisotropic: \\ Profiles derived from \\ PSP observations\\ (\cite{Short2024}) \\ Isotropic: \\ $T_0=(2T_{\perp0}+T_{\parallel0})/3$ \\ with the anisotropic profile} & \makecell[c]{$\hat{B}_{\perp}^2 \propto R^{0.96}$ \\ (\cite{Huang2023c})} \\ \tableline
%
\enddata
\end{deluxetable*} \label{Table1}

\subsection{Background solar wind models}\label{sec:style}

As plasma emanating from the Sun flows radially outward, it undergoes expansion, and the background plasma parameters evolve with the radial distance. To investigate the radial evolution of {$\gamma_{\rm{max}}/\omega_0$} under various solar-wind conditions, we set three radial background models in the range from $R_0$ ($ = 1.02R_\odot$) to $30R_0$. In each case, we prescribe the magnetic field $B_{0}$, mass density $\rho_{0}$, temperature (either $T_0$ or $T_{\parallel0}$ and $T_{\perp0}$), and $\hat{B}_{\perp}$ as functions of the normalized radial distance $\hat{R}=R/R_{0}$ (where $R_{0} = 1.02 R_{\odot}$ denotes the lower coronal boundary), based on theoretical assumptions and, where appropriate, results from previous studies. We assume that $\omega_0$ does not depend on $R$ although some previous studies consider it \citep[e.g.,][]{TeneraniVelli2013}. In other words, the present setup does not explicitly incorporate the effect of solar-wind acceleration. We also assume that these parameters depend only on $R$. Table \ref{Table1} shows three expansion cases considered in this study. 

\subsubsection{Case 1: Spherically symmetric adiabatic expansion with constant wind speed}
In case~1, both the isotropic and anisotropic models assume adiabatic evolution; we examine how the difference between an MHD-type adiabatic assumption and a CGL-type (double-adiabatic) assumption affects the radial evolution of $\gamma_{\rm{max}}/\omega_0$. 

In terms of the background magnetic field $B_{0}$ and the background mass density $\rho_{0}$ in case~1, we assume spherical expansion and a steady solar wind ($\partial/\partial t =0$, ${v}_r=\mathrm{const}$). Then $\nabla\!\cdot\!\mathbf{B}=0$ and mass conservation equation lead to
\begin{equation}
B_0 \propto R^{-2}, \  \rho_0 \propto R^{-2},
\end{equation}
respectively \citep{GrappinVelliMangeney1993, GrappinVelli1996, Matteini2024}. This applies to both the isotropic and anisotropic formulations.

\begin{deluxetable*}{ccccc}
\tablewidth{0pt}
\tabletypesize{\normalsize}
\tablecaption{Parameter ranges in each case \label{tab:description}}
\tablehead{
\colhead{} & \colhead{$\beta(R)$} & \colhead{$\beta_{\parallel}(R)$} & \colhead{$\xi(R)$} & \colhead{$\hat{B}_{\perp}^2(R)$}
}
\startdata
\makecell[c]{Case 1 \\ (Spherically symmetric adiabatic expansion \\ with constant wind speed)} & \makecell[c]{0.001-0.01, \\ 0.01-0.1, \\ 0.1-1} & \makecell[c]{0.00014-0.13, \\ 0.0014-1.3, \\ 0.014-13} & \makecell[c]{10-0.01} & \makecell[c]{0.01-0.3 (Fig. \ref{Fig1}), \\ 0.001-0.03 (Fig. \ref{Fig2})} \\ \hline
\makecell[c]{Case 2 \\ (Multi-source observation- and \\ model-constrained expansion)} & \makecell[c]{0.003-4} & \makecell[c]{0.001-6} & \makecell[c]{14-1.5} & \makecell[c]{0.01-0.26 (Fig. \ref{Fig3}), \\ 0.001-0.026  (Fig. \ref{Fig3})} \\ \hline
\makecell[c]{Case 3 \\  (PSP-constrained expansion)} & \makecell[c]{0.02-1.4} & \makecell[c]{0.002-2} & \makecell[c]{15-0.5} & \makecell[c]{0.01-0.26  (Fig. \ref{Fig4}), \\ 0.001-0.026  (Fig. \ref{Fig4})} \\ \hline
\enddata
\end{deluxetable*} \label{Tabel2}

In terms of the temperature anisotropy, we apply the double-adiabatic closures such as equations of (\ref{M10}) and (\ref{M11}) to the background. Then we get
\begin{equation}
\frac{T_{\parallel0} B_{0}^{2}}{\rho_{0}^{2}} = \mathrm{const}, \quad
\frac{T_{\perp0}}{B_{0}} = \mathrm{const}.
\end{equation}
By combining this with $B_0 \propto R^{-2}, \ \rho_0 \propto R^{-2}$, we can obtain the radial evolution of $T_{\parallel0}$ and $T_{\perp0}$, such as:
\begin{equation}
T_{\parallel0} = \mathrm{const}, \quad T_{\perp0} \propto R^{-2}.
\end{equation}
In the isotropic case, we assume $n=5/3$ and use the adiabatic equation of state
\begin{equation}
T_{0}\,\rho_{0}^{-2/3} = \mathrm{const},
\end{equation}
to obtain the radial evolution of the isotropic temperature. By combining this with $\rho_{0} \propto R^{-2}$, the radial evolution of $T_{0}$ is 
\begin{equation}
T_0 \propto R^{-4/3}.
\end{equation}

In terms of the mother wave amplitude $B_{\perp}$, we adopt 
\begin{equation}
B_{\perp} \propto R^{-1.5},
\end{equation}
based on Wentzel–Kramers–Brillouin (WKB) theory \citep[e.g.,][]{Whang1973}. By combining this with $B_0 \propto R^{-2}$ we can obtain
\begin{equation} \label{WKBRadialEvolution}
\hat{B}_{\perp}^2 \propto R.
\end{equation}
These radial evolutions follow those expected from the Expanding Box Model (EBM) \citep{GrappinVelliMangeney1993, GrappinVelli1996, Matteini2024}. 
Although observations indicate that the solar wind undergoes non-adiabatic expansion \citep{Marsch1982, Marsch1983, Hellinger2011, Hellinger2013},
several numerical studies have investigated PDI while partly assuming adiabatic expansion for investigating the effects of expansion on the wave propagation and the solar wind dynamics \citep[e.g.,][]{TeneraniVelli2013, Reville2018}. In the expanding box model (EBM), it is well known that in the low-$\beta$ regime,
the maximum growth rate scales with heliocentric distance as
$\gamma_{\rm{\max}}/\omega_0 \propto R^{1/3}$ \citep[e.g.,][]{TeneraniVelli2013, Reville2018}.
Although adiabatic expansion is the idealized expansion case, analyzing the radial evolution of PDI under this simplified assumption provides a useful benchmark and enables comparison with previous work.
\subsubsection{Case 2: Multi-source observation- and model-constrained expansion}\label{Case2}
In case~2, we adopt (i) radial profiles of the background magnetic field and density
derived from Metis and PSP observations by \cite{Telloni2021}, (ii) a radial profile of the temperature anisotropy obtained from simulations by \cite{Meng2015}, and (iii) a radial profile of the magnetic-fluctuation amplitude inferred from PSP measurements by \cite{Huang2023c}. Specifically, following \citet{Huang2023c}, who analyzed
PSP observations over $0.06$--$0.3~\mathrm{au}$, we prescribe
\begin{equation}
\hat{B}_{\perp}^2 \propto R^{0.96}.
\end{equation}
This radial evolution is almost equal to the equation (\ref{WKBRadialEvolution}), which is also consistent with results from AWSoM simulation \citep{vanderHolst2022}.
To our knowledge, no observationally derived radial profile of temperature anisotropy within $\sim10R_{\odot}$ is
currently available; therefore, we use the AWSoM simulation results reported by \cite{Meng2015}.
The goal of this case is to quantify how strongly temperature anisotropy can influence PDI under an observation- and model- constrained solar-wind background, but note that this case does not include the $B_\theta$ and $B_\phi$ components in the background magnetic field.
\subsubsection{Case 3: PSP-constrained expansion}\label{Case3}

In case 3, we adopt (i) the PSP-based radial profile of $(\beta_{\parallel},\,\xi)$ based on \cite{Short2024} and (ii) a radial profile of the magnetic-fluctuation amplitude inferred from PSP measurements by \cite{Huang2023c}. The goal of this scenario is also to quantify how strongly temperature anisotropy can influence PDI under the more observationally constrained solar-wind background, which includes the Parker spiral in the background magnetic field. In \autoref{sec:Profiles_of_Previous_studies}, we show the detail of the profiles in this case.

\subsection{\textbf{Boundary values and parameter ranges}}\label{sec:style}


\textbf{To construct the radial profiles of the background plasma parameters, we prescribe either lower-boundary values or fitted radial profiles, depending on the case. Table~\ref{Tabel2} summarizes the resulting ranges of the plasma parameters used to calculate the dispersion relation over $1 \le R/R_0 \le 30$ in each case. For case~1, we impose three lower boundary values of the plasma beta,
$\beta(R_0)=10^{-1},\,10^{-2},\,10^{-3}$ \citep{Gary2001}. We also set the temperature anisotropy to $\xi(R_0)\equiv T_{\perp0}/T_{\parallel0}=10$
\citep{Telloni2007, Cranmer2008, Meng2015, Cranmer2020}. 
For cases~2 and~3, the radial profiles of $\beta_{\parallel}$ and $\xi$ are determined by the profiles described in Section~\ref{Case2}, \ref{Case3}, and Appendix~\ref{sec:Profiles_of_Previous_studies}, rather than by imposing simple lower boundary values. 
In terms of the boundary values of the normalized mother wave amplitude at the corona, we adopt $\hat{B}_{\perp}^{\,2}(R_0)=10^{-2}$ and $\hat{B}_{\perp}^{\,2}(R_0)=10^{-3}$ in all cases
\citep{Reville2018, Hahn2022, vanderHolst2022}.}





\section{Results} \label{sec:floats}

\begin{figure*}[ht!]
\centering
\includegraphics[width=\textwidth,height=0.7\textheight,keepaspectratio]{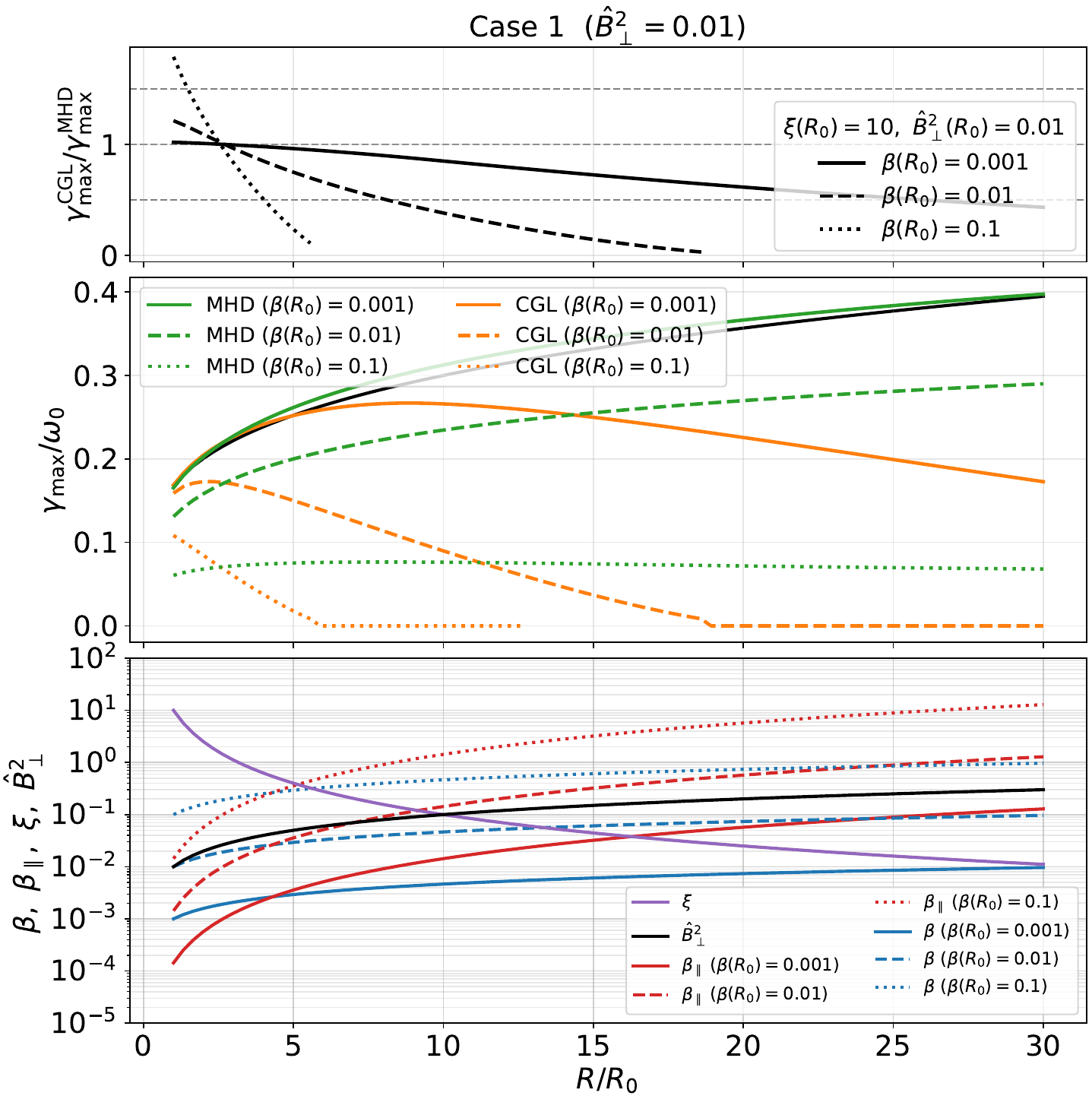}
\caption{\label{fig:8}Radial evolution of the maximum growth rate and the corresponding background plasma parameters for case~1 with $\beta(R=R_0)=0.001,\,0.01,\,0.1$, $\xi(R_0)=10$, and $\hat{B}_{\perp}^{2}(R_0)=0.01$.
The upper panel shows the ratio of the maximum growth rate in the CGL model to that in the ideal MHD model, $\gamma_{\max}^{\mathrm{CGL}}/\gamma_{\max}^{\mathrm{MHD}}$. Three gray horizontal dashed lines indicate $0.5, 1, 1.5$ respectively.
The solid, dashed, and dotted lines correspond to $\beta(R_0)=0.001$, $0.01$, and $0.1$,
respectively. 
The middle panel presents the radial evolution of the maximum growth rate in both models: the green curves denote the ideal-MHD results, whereas the orange curves denote the CGL results. The solid black line is an auxiliary line corresponding to $\gamma_{\max}/\omega_{0} \propto R^{1/4}$. Green and orange solid, dashed, and dotted lines correspond to $\beta(R_0)=0.001$, $0.01$, and $0.1$, respectively.
The lower panel shows the radial evolution of the local background parameters used to evaluate the maximum growth rate. The purple curve indicates the temperature anisotropy $\xi \equiv T_{\perp0}/T_{\parallel0}$ (here $\xi(R_{0})=10$), and the black curve indicates $\hat{B}_{\perp}^{2}$ (here $\hat{B}_{\perp}^{2}(R_{0})=0.01$); these two profiles are common to both the ideal-MHD and CGL models. The red and blue curves represent $\beta_{\parallel}$ and $\beta$, respectively, and the solid/dashed/dotted line styles follow the same convention as in the middle panel.
            }\label{Fig1}
\end{figure*}

\begin{figure*}[ht!]
\centering
\includegraphics[width=\textwidth,height=0.7\textheight,keepaspectratio]{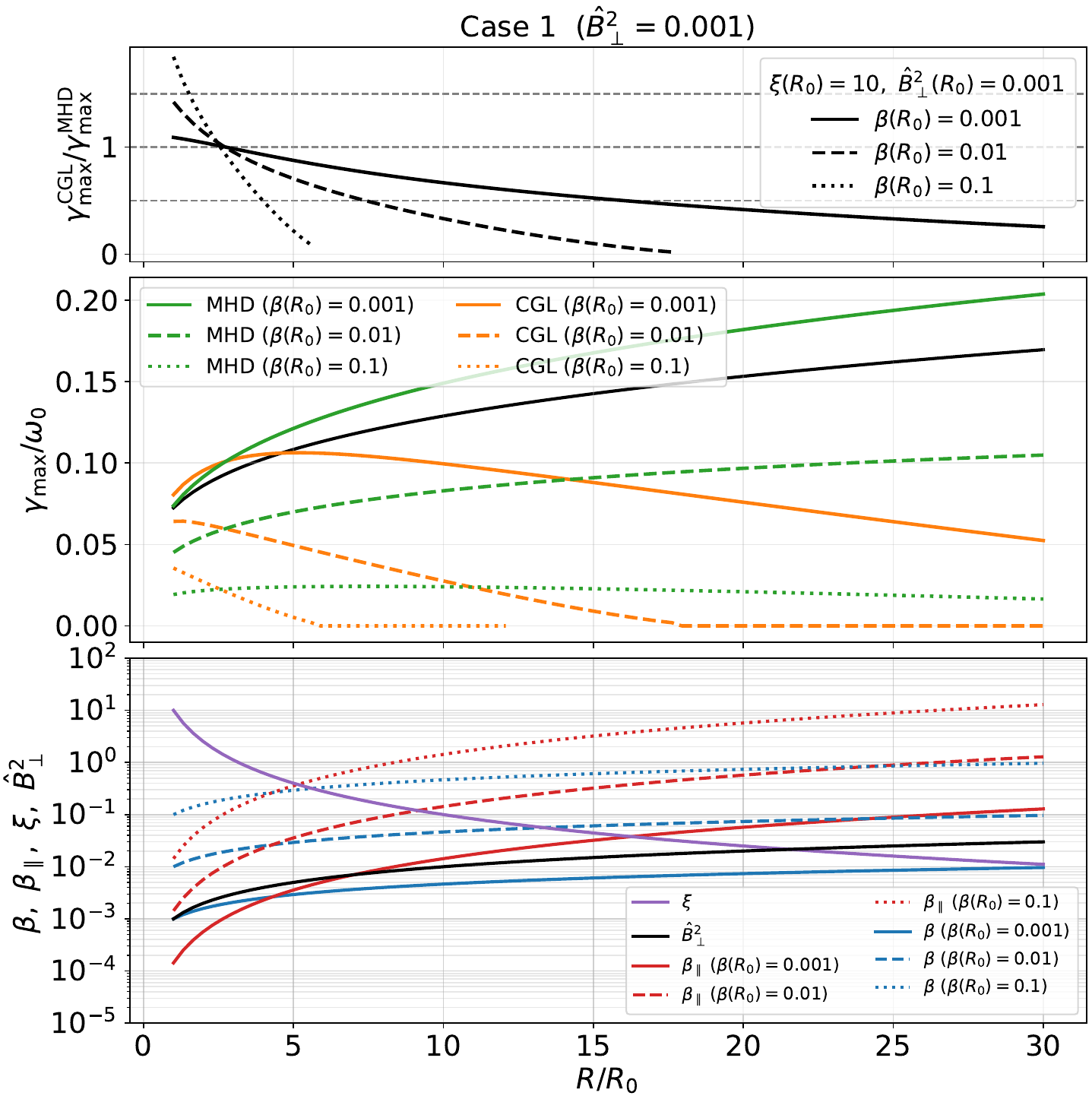}
\caption{\label{fig:8}Radial evolution of the maximum growth rate and the corresponding background plasma parameters for case~1 with $\beta(R_0)=0.001,\,0.01,\,0.1,\,$ $\xi(R_0)=10$, and $\hat{B}_{\perp}^{2}(R_0)=0.001$. The figure format is identical to that of Figure~\ref{Fig1}.}\label{Fig2}
\end{figure*}

Figures~\ref{Fig1} and \ref{Fig2} show the radial evolution of the PDI maximum growth rate in case~1 for two values of the mother-wave amplitude, $\hat{B}_{\perp}^{2}(R_{0}) = 0.01$ and $0.001$, respectively. In each figure, the upper panel shows the ratio $\gamma_{\max}^{\mathrm{CGL}}/\gamma_{\max}^{\mathrm{MHD}}$, the middle panel shows the radial evolution of $\gamma_{\max}/\omega_{0}$ in the two models, and the lower panel shows the corresponding radial profiles of the background parameters used in the calculation.

From the results shown in the upper panels of both Figures~\ref{Fig1} and~\ref{Fig2}, we find that, in the near-Sun region ($R<2R_{0}$), temperature anisotropy with $T_\perp > T_{\parallel}$ increases the PDI maximum growth rate, whereas at larger distances ($R<2R_{0}$) temperature anisotropy with $T_{\parallel}>T_{\perp}$ suppresses it. This trend is consistently confirmed for all three beta cases ($\beta(R_0)=0.1,\,0.01,\,0.001$) and for both amplitude cases ($\hat{B}_{\perp}^{2}(R_0)=0.01,\,0.001$), and it becomes more pronounced as $\beta$ increases. Because the absolute value of the maximum growth rate is smaller at higher $\beta$, the effect of temperature anisotropy appears relatively larger.

For the MHD results, the radial evolution of the maximum growth rate in the middle panels of Figures~\ref{Fig1} and~\ref{Fig2} shows that, for the two low-$\beta$ cases ($\beta(R_0)=0.001$ and $0.01$), the maximum growth rate tends to increase with $R$. In the lowest-$\beta$ case of $\hat{B}_{\perp}^{2}(R_0)=0.01$, the growth rate approximately follows the black solid guideline $\propto R^{1/4}$. Previous studies \citep[e.g.,][]{JayantiHollweg1993} estimated, based on analytical expressions, that under spherically symmetric adiabatic expansion, the maximum growth rate scales as $\propto R^{1/3}$, whereas our results exhibit a slightly different scaling. One plausible reason for this discrepancy is that $\hat{B}_{\perp}^{2}$ is not sufficiently small for the small-amplitude assumption to be strictly valid. Indeed, the smaller-amplitude case (Figure~\ref{Fig2}) is closer to the $R^{1/3}$ scaling than the case of Figure~\ref{Fig1}.

Regarding comparisons with previous studies that examined the radial evolution of the maximum growth rate, the MHD results under a radial evolution of EBM are broadly consistent with earlier works adopting similar assumptions. The tendency for the maximum growth rate to increase with $R$ as a power law in the low-$\beta$ regime agrees with the results of \cite{Reville2018}. The weak decrease of the maximum growth rate with $R$ for the $\beta(R_0)=0.1$ case is consistent with the results of \cite{TeneraniVelli2013}, who explicitly included the $R$-dependence of the pump frequency, as well as with those of \cite{DelZanna2015}, who did not include it (noting, however, that \cite{DelZanna2015} used a somewhat different value, $\hat{B}_{\perp}^{2}=0.04$ at the inner boundary).

For the CGL results, the middle panels of Figures~\ref{Fig1} and~\ref{Fig2} indicate that, for all three beta cases, the maximum growth rate tends to decrease with $R$.
This implies that, within an EBM-CGL framework, the maximum growth rate decays with heliocentric distance, in contrast to the adiabatic EBM-MHD case where it tends to increase with $R$. 

\begin{figure*}[ht!]
\centering
\includegraphics[width=\textwidth,height=0.7\textheight,keepaspectratio]{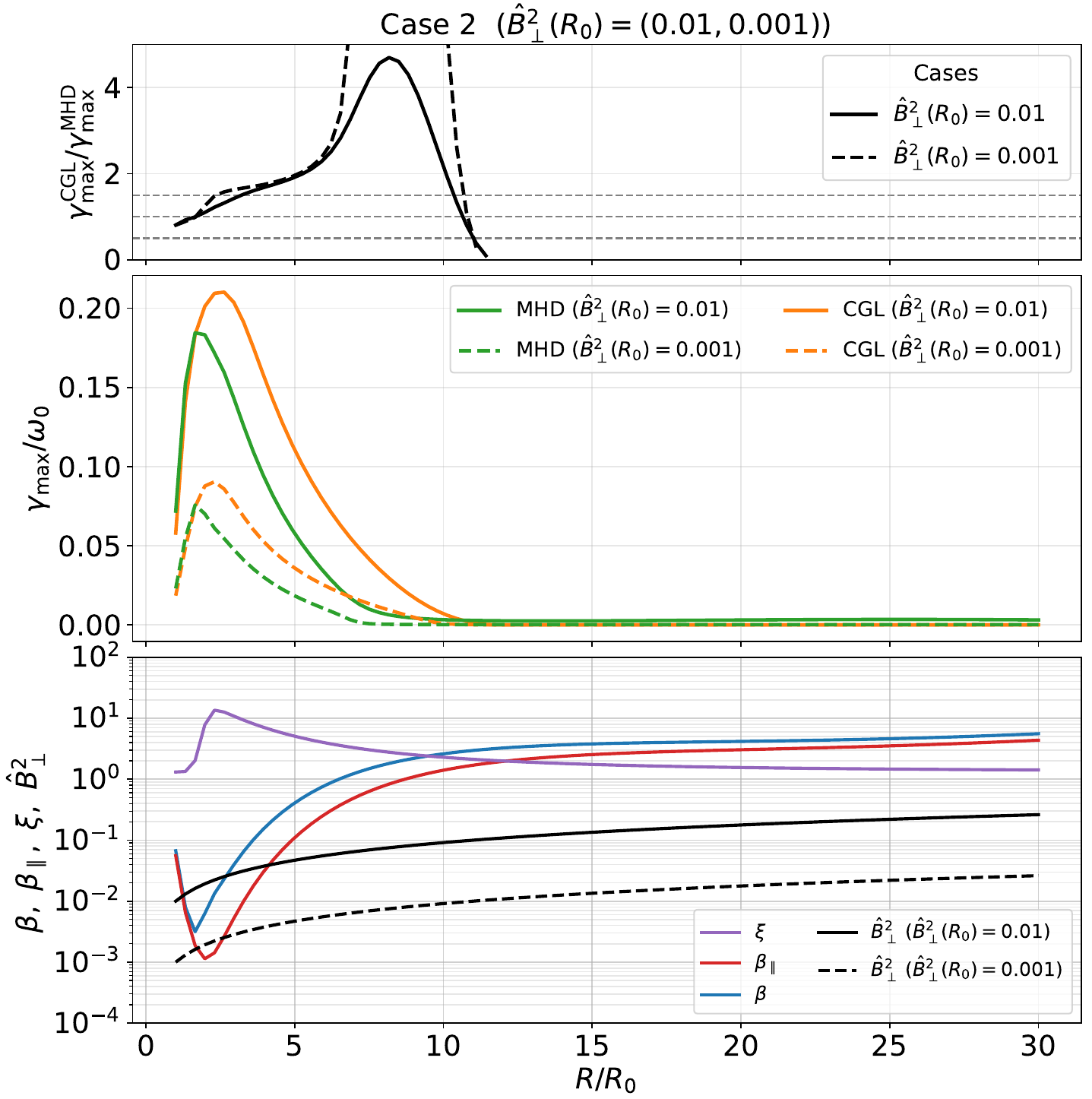}
\caption{\label{fig:8}Radial evolution of the maximum growth rate and the background plasma parameters for case~2 with $\hat{B}_{\perp}^{2}(R_0)=0.01$ and $0.001$.
            The upper panel shows the ratio of the maximum growth rate in the CGL model to that in the ideal MHD model, $\gamma_{\max}^{\mathrm{CGL}}/\gamma_{\max}^{\mathrm{MHD}}$. Three gray horizontal dashed lines indicate $0.5, 1, 1.5$ respectively. The solid and dashed lines correspond to $\hat{B}_{\perp}^{2}(R_0)=0.01$ and $0.001$,
respectively.
The middle panel presents the radial evolution of the maximum growth rate in both models: the green curves denote the ideal-MHD results, whereas the orange curves denote the CGL results. 
The lower panel shows the radial evolution of the local background parameters used to evaluate the maximum growth rate. The purple curve indicates the temperature anisotropy $\xi \equiv T_{\perp0}/T_{\parallel0}$. The black curve indicates $\hat{B}_{\perp}^{2}$ (here $\hat{B}_{\perp}^{2}(R_{0})=0.01, 0.001$); these two profiles are common to both the ideal-MHD and CGL models. The red and blue curves represent $\beta_{\parallel}$ and $\beta$, respectively.
            }\label{Fig3}
\end{figure*}

\begin{figure*}[ht!]
\centering
\includegraphics[width=\textwidth,height=0.7\textheight,keepaspectratio]{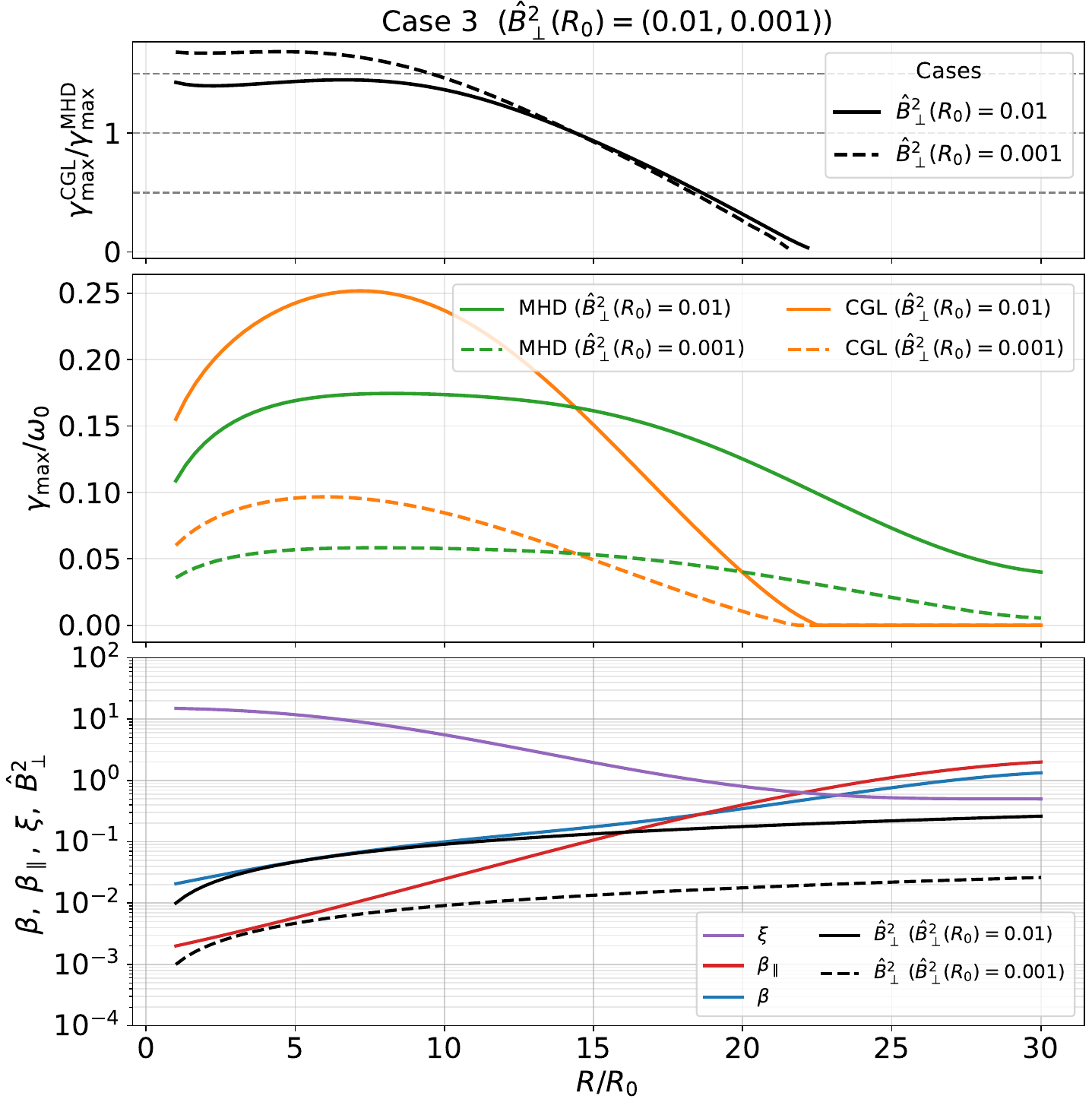}
\caption{\label{fig:8}Radial evolution of the maximum growth rate and the background plasma parameters for case~3 with $\hat{B}_{\perp}^{2}(R_0)=0.01$ and $0.001$ in the common profile of $\beta$ and $T_{\perp0}/T_{\parallel0}$. The figure format is identical to that of Figure~\ref{Fig3}.
            }\label{Fig4}
\end{figure*}

Figure~\ref{Fig3} shows the results for case~2 with $\hat{B}_{\perp}^{2}(R_0)=0.01$ and $0.001$. 
In the upper panel, the ratio of the maximum growth rates reaches $\sim 1.5$ at $R\sim 2R_0$, and then increases up to $\sim 4$ for $\hat{B}_{\perp}^{2}(R_0)=0.01$ and up to $\sim 20$ for $\hat{B}_{\perp}^{2}(R_0)=0.001$. 
Thus, from the upper panel of Figure~\ref{Fig3}, we find that a temperature anisotropy with $T_\perp > T_\parallel$ enhances the maximum growth rate in both cases, $\hat{B}_{\perp}^{2}(R_0)=0.001$ and $0.01$. 
From the middle panel, we find that the PDI is unstable only within $R \lesssim 10R_0$. 
As indicated by the lower panel, this confinement is caused by the rapid increase of $\beta$ and the rapid decrease of $\xi$ associated with solar-wind expansion.
Figure~\ref{Fig4} shows the results for case~3, i.e., the PSP-constrained expansion, again in the same three-panel format as Figure~\ref{Fig3}. 
For the upper panel of Figure~\ref{Fig4}, we confirm again that temperature anisotropy with $T_\perp > T_\parallel$ increases the maximum growth rate by approximately a factor of $\sim 1.5$ within $R < 10R_0$. This result demonstrates that the temperature-anisotropy-driven enhancement of the maximum growth rate remains effective even in a low-$\beta$ regime with $\beta_{\parallel} \lesssim 0.1$. Also, it is clear that, at larger $R$, the maximum PDI growth rate is reduced due to the suppression effects caused by a parallel-dominant temperature anisotropy and by the perturbed temperature anisotropy, which is discussd in Section \ref{sec:highlight1}.

\textbf{
The differences among cases~1--3 can be understood from the radial evolution of $\beta$, $\beta_{\parallel}$, and $\xi$. In case~1, the region where $\gamma_{\max}^{\rm CGL}/\gamma_{\max}^{\rm MHD}>1$ is limited because the double-adiabatic CGL expansion rapidly increases $\beta_{\parallel}$ and decreases $\xi$, while the isotropic-MHD $\beta$ increases more gradually. Thus, the CGL model quickly moves away from the low-$\beta_{\parallel}$ regime where temperature anisotropy with $T_{\perp}>T_{\parallel}$ increases the PDI maximum growth rate. In case~2, the enhancement region where $\gamma_{\max}^{\rm CGL}/\gamma_{\max}^{\rm MHD}>1$ is broader and the peak value of $\gamma_{\max}^{\rm CGL}/\gamma_{\max}^{\rm MHD}$ is larger than in case~1, because temperature anisotropy with $T_{\perp}>T_{\parallel}$ is maintained over a relatively broad range of heliocentric distances. However, the rapid increase in $\beta_{\parallel}$ confines the PDI-unstable region to relatively small heliocentric distances. In case~3, the enhancement region extends over a broader radial interval than in case~2 because both the increase in $\beta_{\parallel}$ and the decrease in $\xi$ are more gradual, although the peak value of the ratio is smaller than in case~2. These comparisons, given the nearly identical radial profiles of $\hat{B}_{\perp}^2$, show that the radial evolution of both $\beta_{\parallel}$ and $\xi$ controls where and how strongly temperature anisotropy enhances or suppresses PDI.
}

Previous studies investigate the radial evolution of the PDI maximum growth rate under conditions closer to the non-adiabatic actual solar wind \citep{ShodaYokoyamaSuzuki2018b, ShodaEtAl2019}. The results of cases 2 and 3 show a faster decrease of the maximum growth rate with increasing $R$ than in \cite{ShodaYokoyamaSuzuki2018b, ShodaEtAl2019}; however, the qualitative feature that the growth rate attains a local maximum---as reported in both \cite{ShodaYokoyamaSuzuki2018b} and \cite{ShodaEtAl2019}---is reproduced. We attribute the faster decline of $\gamma_{\max}$ with increasing $R$ in case 2 to a more rapid increase in $\beta$ than that found in both \cite{ShodaYokoyamaSuzuki2018b} and \cite{ShodaEtAl2019}. A plausible reason for this difference is that those studies considered polar fast wind, which is expected to be less dense and therefore to remain at lower $\beta$ than the low-latitude wind represented by case~2.

\section{Discussion}

\subsection{Parameter scans} \label{ParameterScan}

\begin{figure}[ht!]
\centering
\includegraphics[width=\columnwidth]{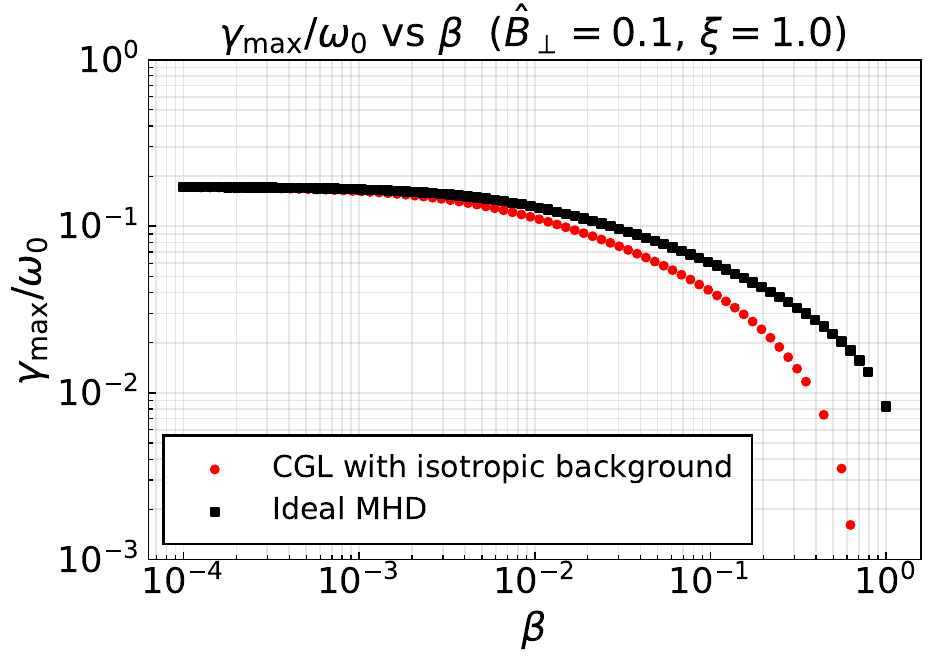}
\caption{\label{fig:4} Maximum growth rate in MHD (black dots) and in CGL (red dots) as a function of $\beta$, evaluated for $\hat{B}_{\perp}=0.1$ and $\xi=1$.
}\label{Fig5}
\end{figure}
\begin{figure}[ht!]
\centering
\includegraphics[width=\columnwidth]{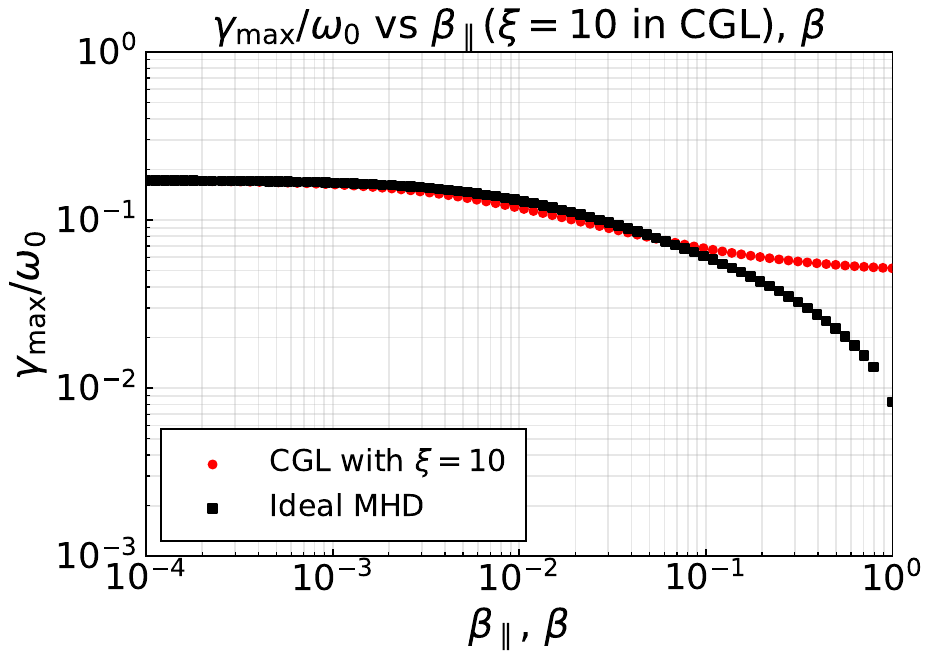}
\caption{Maximum growth rate versus plasma beta in ideal MHD (black dots) and versus parallel plasma beta in CGL with $\xi = 10$ (red dots), both for $\hat{B}_{\perp} = 0.1$. 
}
\label{Fig6_another}
\end{figure}
\begin{figure}[ht!]
\centering
\includegraphics[width=\columnwidth]{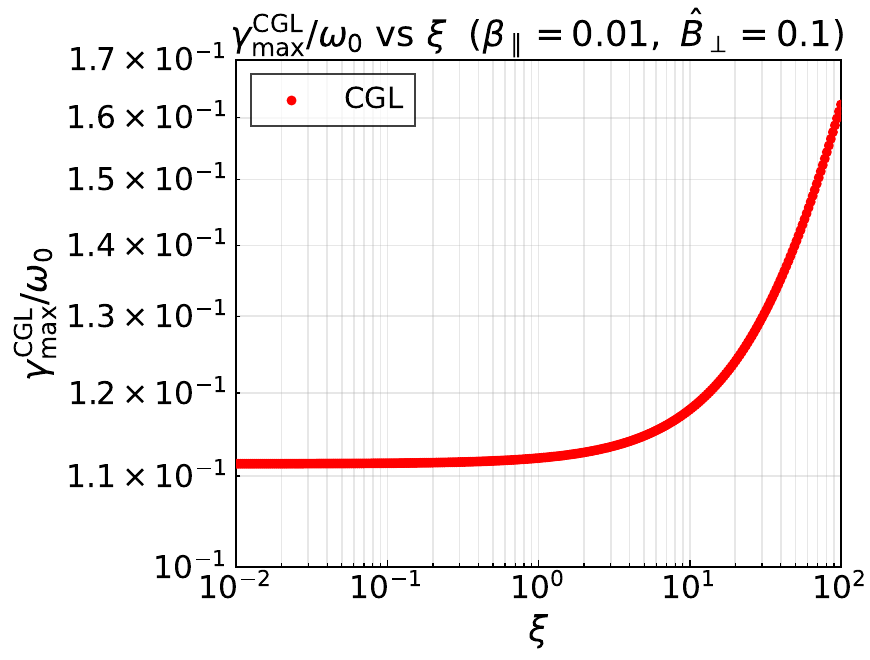}
\caption{\label{fig:4} Maximum growth rate in CGL, $\gamma_{max}^{\rm{CGL}}/\omega_{0}$ (red dots), as a function of $\xi$, evaluated for $\hat{B}_{\perp}=0.1$ and $\beta_{\parallel}=0.01$.
}
\label{Fig6}
\end{figure}

To clarify the physical origin of the $\beta_{\parallel}$-dependence and $\xi$-dependence of the PDI growth rate, we present three parameter-scan figures. Figure~\ref{Fig5} compares $\gamma_{\max}/\omega_{0}$ in ideal MHD and CGL under the same background conditions, with $\xi = 1$ and $\hat{B}_{\perp} = 0.1$ fixed in both models, so that the horizontal axis can be written as the common plasma beta, $\beta (= \beta_{\parallel})$. Figure~\ref{Fig6_another} shows a complementary comparison for an anisotropic background: in the CGL model, $\xi = 10$ and $\hat{B}_{\perp} = 0.1$ are fixed and $\gamma_{\max}/\omega_{0}$ is plotted against $\beta_{\parallel}$, whereas in ideal MHD, $\hat{B}_{\perp} = 0.1$ is fixed and $\gamma_{\max}/\omega_{0}$ is plotted against $\beta$. Figure~\ref{Fig6} isolates the direct $\xi$-dependence of the CGL growth rate by plotting $\gamma_{\max}^{\mathrm{CGL}}/\omega_{0}$ as a function of $\xi$ at fixed $\beta_{\parallel} = 0.01$ and $\hat{B}_{\perp} = 0.1$.

In Figure~\ref{Fig5},
we find that in the low-$\beta$ regime ($\beta \lesssim 10^{-2}$) $\gamma_{\max}/\omega_0$ is nearly identical between MHD and CGL. In contrast, for $\beta \gtrsim 10^{-2}$, the MHD maximum growth rate becomes systematically larger than the CGL value. This difference reflects the effect of perturbed temperature anisotropy, which is discussed in Section~\ref{sec:highlight1}.

Figure~\ref{Fig6_another} illustrates how a temperature anisotropy with $T_\perp > T_\parallel$ modifies the overall $\beta$-dependence of the PDI growth rate in CGL relative to that in ideal-MHD. As shown in Figure~\ref{Fig6_another}, the two models are again similar at sufficiently low $\beta$, whereas toward the higher-$\beta$ side of the scan, the PDI growth rate in CGL remains systematically larger than the ideal-MHD value when $\xi = 10$ is imposed. The results shown in Figures~\ref{Fig5} and \ref{Fig6_another} indicate that, in the low-$\beta$ regime, the dependence of the maximum growth rate on $\beta$ in ideal MHD is essentially identical to its dependence on $\beta_{\parallel}$ in CGL, even when $\xi = 10$ is imposed.

Figure~\ref{Fig6} shows the direct $\xi$-dependence of the PDI growth rate in CGL at fixed $\beta_{\parallel} = 0.01$ and $\hat{B}_{\perp} = 0.1$. As shown in this figure, $\gamma_{\max}^{\mathrm{CGL}}/\omega_{0}$ varies only weakly over a broad range of $\xi$, and a more pronounced increase appears only at sufficiently large $\xi$.

From these results, we interpret the case~1 behavior as follows. In case~1, different adiabatic assumptions were adopted for MHD and CGL, which led to distinct radial scalings: $\beta \propto R^{2/3}$ in MHD, whereas $\beta_{\parallel} \propto R^{2}$ in CGL. The increasing difference between $\beta$ and $\beta_{\parallel}$ with $R$ explains the opposite radial behavior of $\gamma_{\max}$ in MHD and CGL.

\subsection{The effect of the perturbed temperature anisotropy} \label{sec:highlight1}



\begin{figure*}[ht!]
\centering
\includegraphics[width=\textwidth,height=0.7\textheight,keepaspectratio]{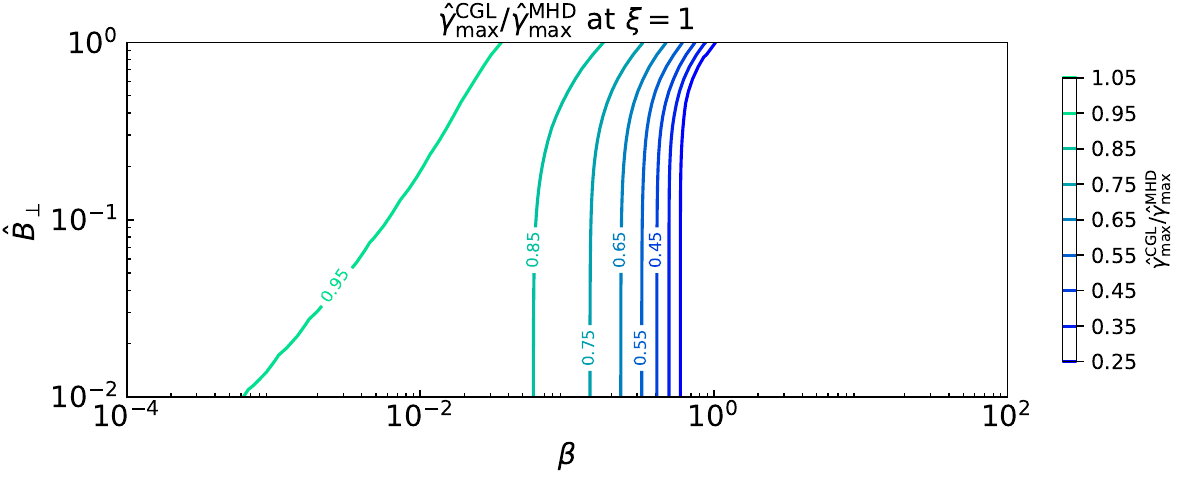}
\caption{\label{fig:8}This figure presents a contour plot of $\gamma_{\max}^{\rm CGL} / \gamma_{\max}^{\rm MHD}$ (contour interval $\Delta=0.1$) for $\xi=1$.
The vertical axis shows $\hat{B}_{\perp}$ over the range $[0.01,1]$, and the horizontal axis shows $\beta$ over the range $[10^{-4},10^{2}]$.
}\label{Fig7}
\end{figure*}



The differences between the maximum growth rates obtained from the isotropic and anisotropic dispersion relations arise from both the background temperature anisotropy and the anisotropy of the perturbations ($\delta T_{\perp} \neq \delta T_{\parallel}$).
Even if we impose a background temperature anisotropy of unity in the CGL-based dispersion relation, (\ref{M12}), the resulting dispersion relation remains completely different from the isotropic (ideal-MHD) dispersion relation, (\ref{M6}). The reason why the dispersion relations differ even under identical background conditions is that the pressure perturbations---i.e., the linearized pressure tensor $\delta{P}$---are not actually the same in the ideal-MHD and CGL models. In other words, because the two linearized sets of governing equations are not equivalent, the dispersion relations derived from them are also different.
However, it can be shown that the two linearized systems become identical if, in addition to setting $\xi = 1$, we further impose $\delta T_{\parallel} = \delta T_{\perp}$, provided that the polytropic index on the MHD side is assumed to be $5/3$ (\autoref{sec:eq_mhd_cgl_equivalence}).

Therefore, allowing $\delta T_{\parallel}$ and $\delta T_{\perp}$ to differ introduces the difference between the two dispersion relations under the same background.
To the best of our knowledge, this study is the first to quantitatively demonstrate the impact of perturbed temperature anisotropy by taking the difference between the ideal-MHD and CGL dispersion relations of the parametric instability.



Figure~\ref{Fig7} shows the ratio $\gamma_{\max}^{\mathrm{CGL}}/\gamma_{\max}^{\mathrm{MHD}}$ in the $(\beta,\hat{B}_{\perp})$ plane for $\xi = 1$. This figure is intended to visualize the impact of perturbed temperature anisotropy by comparing the maximum growth rates in CGL and ideal MHD under the same background anisotropy. As shown in Figure~\ref{Fig7}, the maximum growth rate is generally larger in the ideal-MHD model than in the CGL model. This implies that the perturbed temperature anisotropy (i.e., $\delta T_{\perp} \neq \delta T_{\parallel}$) reduces the maximum growth rate. The magnitude of the difference tends to increase with higher $\hat{B}_{\perp}$ and $\beta$, with the effect being particularly pronounced for $\beta \gtrsim 0.1$.

\subsection{Limitations of this study} \label{sec:cite}
Our study is based on solving linear dispersion relations and therefore does not provide a fully self-consistent description of the coupled evolution of the background plasma and the waves. We quantify how prescribed background properties modify the wave dynamics, whereas the reciprocal feedback of the waves on the background is not captured. Consequently, dedicated simulations that evolve both directions of the coupling are required to assess whether, and to what extent, temperature anisotropy regulates PDI. 

Since PDI can enhance the generation of counter-propagating Alfv\'en waves, it is expected to reduce cross helicity. Since temperature anisotropy with $T_\perp > T_\parallel$ promotes the PDI and increases the generation of counter-propagating waves, the resulting reduction in cross helicity may weaken the helicity barrier, thereby suppressing anisotropic heating. In addition, the relationship between PDI and stochastic heating has been discussed in the literature \citep{Comisel2019}. At present, it remains unclear whether signatures of temperature-anisotropy-enhanced PDI can be identified in the actual solar wind. Further validation from both numerical simulations and in-situ observations is therefore required.

A variety of physical processes not included here can also modify the PDI maximum growth rate, including kinetic effects (e.g., collisionless damping such as Landau damping), large-scale expansion and acceleration, and suppression by background turbulence. In reality, these effects operate simultaneously, making it difficult to quantify the net effect of the PDI. Because these processes act to reduce the PDI growth rate, it is plausible that the effective region where PDI occurs is more restricted than the range inferred from our case~3. Moreover, these effects have often been studied in isolation; a natural next step is to assess PDI in models that integrate multiple physical processes in a unified framework.

The expansion cases considered in this work are somewhat idealized, and there remains room to construct more unideal, observationally constrained expansion cases. In particular, the radial evolution of $\xi$ in the innermost heliosphere is not well captured owing to limited observational coverage. We anticipate that improved empirical constraints on the radial evolution of temperature anisotropy will become feasible as the availability of in-situ measurements continues to increase. For example, observationally based radial trends classified by solar-wind speed and heliographic latitude, and ideally also by source region, would provide valuable constraints.

\section{Conclusion and Summary} \label{sec:highlight}

We investigated how the temperature anisotropy modifies the onset and radial evolution of the
Alfv\'en wave parametric decay instability (PDI) in the near-Sun solar wind.
We prescribed expanding background
profiles and computed the normalized maximum growth rate,
$\gamma_{\max}/\omega_{0}$, by directly solving two linear dispersion relations:
(i) the ideal-MHD dispersion relation of \citet{Goldstein1978,Derby1978} and
(ii) the CGL-based dispersion relation of \citet{Tenerani2017}.
To consider the role of expansion, we considered multiple expansion cases over
$1 \le R/R_0 \le 30$ and emphasized three representative cases in the main text:
case~1 (spherically symmetric adiabatic expansion with constant wind speed), case~2 (Multi-source observation- and model-constrained expansion) and case~3 (PSP-constrained expansion).

Our main findings are summarized as follows.

\begin{enumerate}
\item \textit{Background temperature anisotropy can significantly enhance PDI even in low-$\beta$
near-Sun conditions.}
In case~3, where $\beta_{\parallel} \lesssim 0.1$ in the inner heliosphere, the CGL model yields a maximum
growth rate larger than the ideal-MHD value by a factor of $\sim 1.5$ over $R/R_0\sim 1$--$10$ when the background anisotropy is perpendicular-temperature dominated. 
\textbf{This enhancement is caused by the radial evolution of both $\beta_{\parallel}$ and $\xi$: unlike case~1, where the double-adiabatic expansion rapidly increases $\beta_{\parallel}$ and decreases $\xi$, cases~2 and~3 maintain $T_{\perp}>T_{\parallel}$ over a relatively broad range of heliocentric distances. In particular, case~3 keeps $\beta_{\parallel}$ relatively low while $\xi>1$ persists, allowing $\gamma_{\max}^{\rm CGL}/\gamma_{\max}^{\rm MHD}>1$ to extend over a broader radial interval.}
Moreover, in case~3, temperature anisotropy with $T_{\parallel} > T_{\perp}$ reduces the maximum growth rate, thereby limiting the radial extent of the PDI-unstable region to $R \lesssim 30 R_{0}$. Other effects---such as solar-wind expansion, background turbulence, and Landau damping---are not considered in this study, so our estimated radial range for PDI occurrence could be confined to even smaller heliocentric distances.
Our result demonstrates that anisotropy-driven enhancement remains effective in the parameter regime most relevant to PSP observations, and therefore temperature anisotropy should be treated as a non-negligible control parameter when assessing where PDI can \textbf{occur} in the near-Sun wind.

\item \textit{In the adiabatic case, the radial evolution of $\gamma_{\max}/\omega_{0}$ differs qualitatively between the isotropic and anisotropic closures.}
Under the adiabatic assumption, ideal MHD predicts that $\gamma_{\max} / \omega_0$ increases with heliocentric distance
in the low-$\beta$ regime, with an approximate power-law scaling that approaches the classical
${\gamma}_{\max} /\omega_0 \propto R^{1/3}$ behavior in the smaller-amplitude limit.
In contrast, the CGL model does not exhibit a monotonic increase with $R$. Instead, ${\gamma}_{\max} /\omega_0$ can turn over and decrease beyond a certain distance.
\textbf{The parameter scans in Section~\ref{ParameterScan} suggest that in the low-$\beta$ regime, the PDI growth rates in both models are nearly identical even when $\xi=10$ is imposed, whereas the direct $\xi$-dependence itself remains weak over a broad range. Therefore, because} the adiabatic closures imply different radial scalings, $\beta \propto R^{2/3}$ in MHD and $\beta_{\parallel}\propto R^{2}$ in CGL, the more rapid increase of $\beta_{\parallel}$ in the CGL closure provides a plausible explanation for the different radial behaviors found in case~1.

\item \textit{Perturbed temperature anisotropy reduces ${\gamma}_{\max} /\omega_0$.}
Even when the background anisotropy is set to unity ($\xi=1$), \textbf{the CGL-based dispersion relation} does not
reduce to the ideal-MHD form because the linearized pressure responses differ unless one further imposes
$\delta T_{\perp}=\delta T_{\parallel}$ (for $n=5/3$).
By comparing ${\gamma}_{\max} /\omega_0$ from CGL at $\xi=1$ with the ideal-MHD result, we quantified the 
impact of the perturbed temperature anisotropy.
The resulting maps show that ${\gamma}_{\max} /\omega_0$ is generally smaller in CGL than in MHD, and the
difference grows with increasing $\hat{B}_{\perp}$ and $\beta$.
This implies that allowing $\delta T_{\perp}\neq \delta T_{\parallel}$ tends to suppress PDI.
\end{enumerate}

These results have direct implications for PDI-based interpretations and models of the near-Sun solar wind.
Because PDI produces compressive fluctuations and counter-propagating Alfv\'enic components, any
temperature-anisotropy-driven modulation of ${\gamma}_{\max} /\omega_0$ can influence the generation of density fluctuations,
the effective reflection rate, and the subsequent development of Alfv\'en-wave turbulence.
Our findings therefore motivate incorporating anisotropic thermodynamics when estimating PDI activity
in inner-heliosphere applications, including global simulations that use PDI as an
effective mechanism for controlling turbulence.

Finally, several caveats and future directions follow naturally from this work.
Our calculations are based on linear dispersion relations with prescribed backgrounds; thus, the
two-way feedback between wave evolution (including PDI-driven compressive and backward components)
and the background thermodynamics is not captured.
Moreover, additional processes known to reduce PDI growth in the solar wind---such as expansion and
acceleration, turbulence and kinetic damping---were not included
and may further confine the effective PDI operating region closer to the Sun than suggested by our case~3.
A key next step is to test temperature-anisotropy-modified PDI using simulations that evolve both the wave dynamics
and the anisotropic background self-consistently.

\begin{acknowledgments}
This work was financially supported by the WISE Program for Sustainability in the Dynamic Earth (SyDE), Tohoku University. The work of YK is supported by JSPS KAKENHI grant 25K00975. MS is supported by JSPS KAKENHI Grant Number JP24K00688. HS thanks K. Isono, K. Saito and W. Ishizaki for the supports on code development.
\end{acknowledgments}

\begin{contribution}

H. Saguchi led the analysis and manuscript preparation. Y. Kawazura, M. Shoda, and Y. Katoh provided valuable discussions and feedback and reviewed the entire manuscript.


\end{contribution}


\appendix

\section{Profiles that we use: Telloni et al. 2021, Meng et al. 2015 and Short et al. 2024} \label{sec:Profiles_of_Previous_studies} 



In case 2, we use the magnetic-field profile made by \citet{Telloni2021}. We constructed the profile over $1$--$30\,R_{0}$ by fitting the data in $\log_{10}$--$\log_{10}$ space with a fourth-order polynomial. The resulting profile gives $B\simeq 1.7\times10^{2}\,\mathrm{nT}$ at $\hat{R}\equiv R/R_{0}=30$. The fitted expression is
\begin{equation}
    \log_{10} B(\hat{R}) = -0.901205(\log_{10} \hat{R})^4 + 3.566625(\log_{10} \hat{R})^3 - 3.448904 (\log_{10} \hat{R})^2- 2.331466 (\log_{10} \hat{R}) + 6.003316.
\end{equation}
We note that some observational studies suggest values closer to $100\,\mathrm{nT}$ at $R\sim 30\,R_{0}$ \citep[e.g.,][]{Perrone2019, Maruca2023}. Although the value differs slightly, the order of magnitude is consistent; therefore, we adopted this fitting profile.
We also use the electron density profile made by \citet{Telloni2021} for deriving the density profile $n(R)$ under the assumption $n_{e} \approx n_{p}$. We constructed the density profile over the range $1$--$30\,R_{0}$, $R_{0}=1.02R_{\odot}$ , using a fitted function based on the fitting data of \citet{Telloni2021}, after adding an additional anchor point of $n=10^{9}\,\mathrm{cm^{-3}}$ at $\hat{R}\equiv R/R_{0}=1$ based on \citet{Gary2001}. We performed the fit in $\log_{10}$--$\log_{10}$ space using a fourth-order polynomial, because \citet{Telloni2021} also describes their fitting as a fourth-order polynomial in logarithmic space. The resulting fitted expression is
\begin{equation}
    \log_{10} n(\hat{R}) = 2.8999(\log_{10}\hat{R})^4 - 11.2618(\log_{10} \hat{R})^3 + 16.7644(\log_{10} \hat{R})^2 - 13.3988 (\log_{10} \hat{R}) + 8.9892.
\end{equation}


In terms of the profile of temperature, we use the temperature-anisotropy profile of \citet{Meng2015} in cases 2. We extracted the data points over the range $1$--$30\,R_{0}$ and constructed a fitted profile from these datasets. To obtain a smooth analytic representation of the radial evolution of the anisotropy, we fitted the data with a ``bump'' model defined in logarithmic radius. Specifically, we modeled $\xi(\hat{R})\equiv T_{\perp}/T_{\parallel}$ as a baseline level plus a localized enhancement that rises and falls with $R$, expressed as the product of two logistic functions in $\ln \hat{R}$:
\begin{equation}
\xi(\hat{R})=A + D\,L_{\rm rise}(\hat{R})\,\bigl[1-L_{\rm fall}(\hat{R})\bigr],
\label{eq:xi_bump}
\end{equation}
where
\begin{equation}
L_{\rm rise}(\hat{R})=\left[1+\exp\left\{-k_{1}\left(\ln \hat{R}-\mu_{1}\right)\right\}\right]^{-1},
\label{eq:xi_rise}
\end{equation}
\begin{equation}
L_{\rm fall}(R)=\left[1+\exp\left\{-k_{2}\left(\ln R-\mu_{2}\right)\right\}\right]^{-1}.
\label{eq:xi_fall}
\end{equation}
Here, $A$ represents the background anisotropy level, $D$ controls the amplitude of the enhancement, and $(k_{1},\mu_{1})$ and $(k_{2},\mu_{2})$ set the steepness and the characteristic locations of the rise and fall, respectively. The parameters were obtained by a bounded nonlinear least-squares fit to the extracted data points, yielding the best-fit values
\begin{equation}
(A,\,D,\,k_{1},\,\mu_{1},\,k_{2},\,\mu_{2})
=
(1.445318,\,50.000000,\,16.603014,\,0.710795,\,1.930799,\,0.321134).
\label{eq:xi_bump_params}
\end{equation}


 To construct a smooth radial profile of the parallel proton temperature, $T_{\parallel}(R)$, we digitized the data points from \citet{Meng2015} within $R_{0}$--$30R_{0}$ and fitted them with a smoothing spline in logarithmic radius. Specifically, we first sorted the extracted pairs $\{(R_i,\,T_{\parallel,i})\}$ in ascending $R$ and introduced the logarithmic coordinate
\begin{equation}
x \equiv \ln\!\left(\frac{R}{R_{\odot}}\right).
\label{eq:x_logR}
\end{equation}
We then modeled $T_{\parallel}$ as a smooth function of $x$ using a cubic smoothing spline,
\begin{equation}
T_{\parallel}(\hat{R}) = S(x),
\label{eq:Tpara_spline}
\end{equation}
where $S(x)$ is obtained by minimizing the standard smoothing-spline functional
\begin{equation}
\sum_{i=1}^{N}\left[T_{\parallel,i}-S(x_i)\right]^2
\;+\;
s\int \left[S''(x)\right]^2\,dx,
\label{eq:smoothing_spline}
\end{equation}
where $x_i=\ln(R_i/R_{\odot})$, $s=10^{11}$. The integral penalizes excessive curvature of the fitted profile. Here, the integration interval $[x_{\min},x_{\max}]$ corresponds to the radial range covered by the extracted dataset (approximately $R\sim 1$--$50\,R_{\odot}$). The smoothing parameter $s$ controls the trade-off between fidelity to the data (first term) and smoothness of the profile (second term). Using the resulting spline $S(x)$, we evaluated $T_{\parallel}(\hat{R})$ on a uniform grid in $\hat{R}$ over $1$--$30\,R_{0}$ for use in our calculations.

In case 3, we use the radial profile of the proton parallel plasma beta and temperature anisotropy of \cite{Short2024} for deriving $\beta_{\parallel}$ and $\xi$ under the assumption $n_{e} \approx n_{p}, \ T_{e} \approx T_{p}$. We introduce smooth analytic functions of
the normalized heliocentric distance $\hat{R}\equiv R/R_0$ ($1\le \hat{R}\le 30$).
The coefficients are chosen so that the functions reproduce the prescribed
values at four anchor radii ($\hat{R}=1,13,20,30$) exactly, while avoiding spurious
oscillations and ensuring positivity. The three anchor radii at $\hat{R}=13$, $20$, and $30$ are from \cite{Short2024}, while an additional anchor point at $\hat{R}=1$ is introduced based on \cite{ShodaYokoyamaSuzuki2018b}, although \cite{ShodaYokoyamaSuzuki2018b} set $\beta\sim0.014$ at $R=1.014R_{\odot}$. For $\beta=0.01$, the mirror-instability threshold \citep[][]{Hellinger2006} is located at $(\beta_{\parallel}, \xi) = (8 \times 10^{-4}, 18)$, so we adopt $(\beta_{\parallel}, \xi) = (1 \times 10^{-3}, 15)$ as the anchor point at $R=R_{0}\sim1.02R_{\odot}$. We note that \cite{Short2024} report the \emph{proton} parallel plasma beta, $\beta_{\parallel p} \equiv 8\pi p_{\parallel p}/B^{2}$; here, for simplicity, we use it as a proxy for the parallel plasma beta.
We model $\beta_{\parallel}(\hat{R})$ by interpolating $\log_{10}\beta_{\parallel}$
with a cubic polynomial,
\begin{equation}
\beta_{\parallel}(\hat{R})=10^{f_{\beta}(\hat{R})},\qquad
f_{\beta}(r)=a \hat{R}^{3}+b \hat{R}^{2}+c \hat{R}+d,
\label{eq:beta_profile}
\end{equation}
where the coefficients are
\begin{align}
a&=-8.71798336\times10^{-5},\nonumber\\
b&=\ \ 2.68031491\times10^{-3},\nonumber\\
c&=\ \ 1.01522939\times10^{-1},\nonumber\\
d&=-3.10411607.
\end{align}
This construction guarantees $\beta_{\parallel}(\hat{R})>0$ for all $\hat{R}$ and reproduces
the anchor values
$\beta_{\parallel}(1)=10^{-3}$,
$\beta_{\parallel}(13)=3\times10^{-2}$,
$\beta_{\parallel}(20)=2\times10^{-1}$, and
$\beta_{\parallel}(30)=1$.

In terms of $\xi$, to obtain a single smooth function that is strictly monotonic decreasing,
we parametrize $\xi(\hat{R})$ as
\begin{equation}
\xi(\hat{R})=\xi_1 \exp\!\left[-I(\hat{R})\right],\qquad \xi_1\equiv \xi(1)=15,
\label{eq:xi_profile_def}
\end{equation}
with
\begin{equation}
\frac{dI}{d\hat{R}}=\left(u \hat{R}^2+v \hat{R}+w\right)^2 \ge 0 .
\label{eq:xi_monotone_constraint}
\end{equation}
Because $d\ln\xi/d\hat{R}=-dI/d\hat{R}\le 0$, Eq.~(\ref{eq:xi_monotone_constraint})
guarantees that $\xi(\hat{R})$ is monotonic decreasing for all $\hat{R}$.
Integrating Eq.~(\ref{eq:xi_monotone_constraint}) from $\hat{R}=1$ yields
\begin{equation}
I(\hat{R})=A_5\left(\hat{R}^5-1\right)+A_4\left(\hat{R}^4-1\right)+A_3\left(\hat{R}^3-1\right)
     +A_2\left(\hat{R}^2-1\right)+A_1\left(\hat{R}-1\right),
\label{eq:xi_profile_poly}
\end{equation}
so that
\begin{equation}
\xi(\hat{R})=15\exp\!\left\{-\left[
A_5(\hat{R}^5-1)+A_4(\hat{R}^4-1)+A_3(\hat{R}^3-1)+A_2(\hat{R}^2-1)+A_1(\hat{R}-1)
\right]\right\}.
\label{eq:xi_profile_final}
\end{equation}
The coefficients used in this work are
\begin{align}
A_5&= 8.503658213699558\times10^{-7},\nonumber\\
A_4&=-5.682864240998027\times10^{-5},\nonumber\\
A_3&= 8.835153670285457\times10^{-4},\nonumber\\
A_2&= 5.1814868111597565\times10^{-3},\nonumber\\
A_1&= 8.836698570459859\times10^{-3}.
\end{align}
These coefficients reproduce the anchor values
$\xi(1)=15$, $\xi(13)=3$, $\xi(20)=0.8$, and $\xi(30)=0.5$ exactly.



\section{Origin of the difference between the ideal MHD and CGL dispersion relations at $\xi = 1$} \label{sec:eq_mhd_cgl_equivalence}
First, we set only the background temperature anisotropy to unity and show that, even in this case, the linearized pressure tensor is not identical.
The pressure tensor in CGL is
\begin{equation}
\boldsymbol{P} =p_{\perp} \boldsymbol{I} + (p_{\parallel} - p_{\perp}) \boldsymbol{b}\boldsymbol{b}.
\label{B1}
\end{equation}
Here, $\mathbf{I}$ is the identity tensor and $\boldsymbol{b}=\boldsymbol{B}/B$ is the unit vector along the local magnetic-field direction. We decompose the variables into a background and small perturbations as
\begin{equation}
    \rho = \rho_{0} + \delta{\rho}, \quad p_{\perp}=p_{\perp0}+\delta p_{\perp}, \quad p_{\parallel}=p_{\parallel0}+\delta p_{\parallel}, \quad \boldsymbol{b}=\boldsymbol{b}_{0}+\delta \boldsymbol{b},
\label{B2}
\end{equation}
where $\boldsymbol{b_{0}}$ includes the background magnetic field $B_{0}$ and the mother wave amplitude $B_{\perp}$.
In this case, the zeroth-order pressure tensor is
\begin{equation}
\boldsymbol{P}_{0}=p_{\perp0} \boldsymbol{I} + (p_{\parallel0}-p_{\perp0})\boldsymbol{b}_{0}\boldsymbol{b}_{0}.
\label{B3}
\end{equation}
The first-order pressure tensor is
\begin{equation}
    \delta{\boldsymbol{P}} = \delta{p_{\perp}} \boldsymbol{I} + (\delta{p_{\parallel}}-\delta{p_{\perp}}) \boldsymbol{b}_{0} \boldsymbol{b}_{0} + (p_{\parallel0}-p_{\perp0})(\boldsymbol{b}_{0}\delta{\boldsymbol{b}}+\delta{\boldsymbol{b}\boldsymbol{b}_{0}}).
\label{B4}
\end{equation}
On the other hand, when the linearization in ideal MHD is assumed to be
\begin{equation}
    \rho = \rho_{0} + \delta{\rho}, \quad p = p_{0} + \delta{p}, \quad \boldsymbol{b} = \boldsymbol{b_{0}}+\delta{\boldsymbol{b}}.
\label{B5}
\end{equation}
The pressure tensor in ideal MHD is 
\begin{equation}
    \boldsymbol{P} = p \boldsymbol{I}.
\label{B6}
\end{equation}
So, the zeroth-order pressure tensor is 
\begin{equation}
    \boldsymbol{P}_{0} = p_{0} \boldsymbol{I},
\label{B7}
\end{equation}
and the first-order pressure tensor is
\begin{equation}
    \delta{\boldsymbol{P}} = \delta{p} \boldsymbol{I}.
\label{B8}
\end{equation}
Therefore, even though we assume $\xi = p_{\perp0}/p_{\parallel0}$ = 1, the pressure in CGL becomes
\begin{equation}
    \delta{\boldsymbol{P}} = \delta{p_{\perp}} \boldsymbol{I} + (\delta{p_{\parallel}}-\delta{p_{\perp}}) \boldsymbol{b}_{0} \boldsymbol{b}_{0} + \cancel{(p_{\parallel0}-p_{\perp0})(\boldsymbol{b}_{0}\delta{\boldsymbol{b}}+\delta{\boldsymbol{b}\boldsymbol{b}_{0}})},
\label{B9}
\end{equation}
not equal to the pressure tensor of ideal MHD, the equation (\ref{B8}) due to $\delta{p}_{\perp} \neq \delta{p}_{\parallel}$. So, roughly speaking, the linearized set of governing equations in CGL is not the same as ideal MHD (though the differential equations for pressure in CGL are not the same as in ideal MHD). The dispersion relation of the parametric instability is derived from a set of linearized equations of governing equations, so the dispersion relation in CGL is not equal to the dispersion relation in ideal MHD even if $\xi=1$.

If we only assume $\xi=1$, three of the governing equations, such as the mass conservation, the momentum equation, and the induction equation in CGL like equations~(\ref{M7})--(\ref{M9}), are equal to three of equations in ideal MHD like equations~(\ref{M1})--(\ref{M3}). So, these three linearized equations in CGL are also equal to three linearized equations in ideal MHD when $\xi=1$.

It should also be emphasized that imposing $\xi=1$ and $\delta p_{\parallel}=\delta p_{\perp}$ is not sufficient for the linearized governing equations to become identical between the ideal-MHD and CGL frameworks. In addition, one must assume a polytropic closure with the polytropic index $n=5/3$; otherwise, the linearized equation sets do not coincide. We demonstrate this point below. Linearized equations of equations~(\ref{M10}), (\ref{M11}) are
\begin{equation}
    \frac{\delta{p_{\perp}}}{p_{\perp}}=\frac{\delta{\rho}}{\rho_{0}}+\frac{\delta{B}}{B_{0}}, \quad
    \frac{\delta{p_{\parallel}}}{p_{\parallel}}=\frac{3\delta{\rho}}{\rho_{0}}-\frac{2\delta{B}}{B_{0}}.
\label{B10}
\end{equation}
Then if we assume $p_{\perp}=p_{\parallel}=p$ and $\delta p_{\parallel}=\delta p_{\perp}=\delta{p}$, equations (\ref{B10}) become like
\begin{equation}
    \frac{\delta{p}}{\delta{\rho}}=\frac{5}{3}\frac{p_{0}}{\rho_{0}}.
\end{equation}
This linearized relation is equivalent to the linearized ideal-MHD closure in equation (\ref{M4}) only when the polytropic index is set to $n=5/3$.

So, if we assume the polytropic index $n=5/3$, \ $p_{\perp}=p_{\parallel}=p$ and $\delta p_{\parallel}=\delta p_{\perp}=\delta{p}$, a set of linearized equations in CGL is equal to a set of linearized equations in ideal MHD; therefore, the dispersion relation in CGL is equal to the dispersion relation in ideal MHD. 
For these reasons, even when we set $n=5/3$ and $\xi=1$, the maximum growth rates obtained from the two dispersion relations do not become identical, because of $\delta p_{\parallel} \neq \delta p_{\perp}$.

\bibliography{sample701}{}

@article{Gary2001,
  author  = {Gary, G. Allen},
  title   = {Plasma Beta above a Solar Active Region: Rethinking the Paradigm},
  journal = {Solar Physics},
  year    = {2001},
  month   = {10},
  volume  = {203},
  number  = {1},
  pages   = {71--86},
  doi     = {10.1023/A:1012722021820}
}

@article{Coello-Guzman2026,
  author        = {Coello-Guzm{\'a}n, Matilde and Pinto, V{\'\i}ctor A. and Navarro, Roberto E. and Moya, Pablo S.},
  title         = {The Effect of Expansion and Instabilities in the Thermodynamic Regulation of the Young Solar Wind Plasma},
  year          = {2026},
  eprint        = {2603.25443},
  archivePrefix = {arXiv},
  primaryClass  = {astro-ph.SR},
  url           = {https://arxiv.org/abs/2603.25443}
}

@article{GonzalezEtAl2026,
  title   = {Characterization of Compressible Fluctuations in Solar Wind Streams Dominated by Balanced and Imbalanced Turbulence: Parker Solar Probe, Solar Orbiter, and Wind Observations},
  author  = {Gonz{\'a}lez, C. A. and Gonzalez, C. and Tenerani, A.},
  journal = {The Astrophysical Journal},
  volume  = {1002},
  number  = {1},
  pages   = {94},
  year    = {2026},
  month   = apr,
  doi     = {10.3847/1538-4357/ae5bb4}
}

@article{Yoon2026,
  author  = {Yoon, Peter H. and Salem, Chadi S. and Lazar, Marian and Martinovi{\'c}, Mihailo M. and Klein, Kristopher G. and L{\'o}pez, Rodrigo A. and Shaaban, Shaaban M. and Seough, Jungjoon and Huang, Jia and Sarfraz, Muhammad and Poedts, Stefaan},
  title   = {Regulation of temperature anisotropy for solar wind protons and alpha particles by collisions and instabilities},
  journal = {Astronomy \& Astrophysics},
  volume  = {705},
  pages   = {A14},
  year    = {2026},
  month   = jan,
  doi     = {10.1051/0004-6361/202557070}
}

@article{ChandranEtAl2025a,
  author  = {Chandran, Benjamin Divakar Giles and Adkins, Toby and Bale, Stuart D. and David, Vincent and Halekas, Jasper and Klein, Kristopher and Meyrand, Romain and Perez, Jean C. and Shoda, Munehito and Squire, Jonathan and Yerger, Evan Lowell},
  title   = {A two-fluid solar-wind model with intermittent Alfv{\'e}nic turbulence},
  journal = {Journal of Plasma Physics},
  volume  = {91},
  number  = {4},
  year    = {2025},
  month   = aug,
  doi     = {10.1017/S0022377825100640}
}

@article{OfmanGaryVinas2002,
  author  = {Ofman, Leon and Gary, S. Peter and Vi{\~n}as, Adolfo F.},
  title   = {Resonant heating and acceleration of ions in coronal holes driven by cyclotron resonant spectra},
  journal = {Journal of Geophysical Research: Space Physics},
  volume  = {107},
  number  = {A12},
  pages   = {1461},
  year    = {2002},
  doi     = {10.1029/2002JA009432}
}

@article{OfmanVinasGary2001,
  author  = {Ofman, Leon and Vi{\~n}as, Adolfo F. and Gary, S. Peter},
  title   = {Constraints on the O$^{5+}$ Anisotropy in the Solar Corona},
  journal = {The Astrophysical Journal Letters},
  volume  = {547},
  pages   = {L175--L178},
  year    = {2001},
  doi     = {10.1086/318900}
}

@article{Johnston2025,
  author  = {Johnston, Zade and Squire, Jonathan and Meyrand, Romain},
  title   = {Unified Phenomenology and Test-Particle Simulations of Ion Heating in Low-$\beta$ Plasmas},
  journal = {Physical Review Letters},
  volume  = {135},
  pages   = {095201},
  year    = {2025}
}

@article{Hellinger2006,
  author  = {Hellinger, Petr and Tr{\'a}vn{\'\i}{\v{c}}ek, Pavel and Kasper, Justin C. and Lazarus, Alan J.},
  title   = {Solar wind proton temperature anisotropy: Linear theory and WIND/SWE observations},
  journal = {Geophysical Research Letters},
  volume  = {33},
  number  = {9},
  pages   = {L09101},
  year    = {2006},
  doi     = {10.1029/2006GL025925}
}

@article{Mallet2026,
  author  = {Mallet, Alfred and Klein, Kristopher G. and Chandran, Benjamin D. G. and Ervin, Tamar and Bowen, Trevor A.},
  title   = {Perpendicular ion heating in turbulence and reconnection: magnetic moment breaking by coherent fluctuations},
  journal = {Journal of Plasma Physics},
  volume  = {92},
  number   = {1},
  pages   = {E13},
  year    = {2026},
  doi     = {10.1017/S0022377825101177}
}

@article{Meng2015,
  author  = {Meng, Xing and van der Holst, Bart and T{\'o}th, G{\'a}bor and Gombosi, T. I.},
  title   = {Alfv{\'e}n wave solar model (AWSoM): proton temperature anisotropy and solar wind acceleration},
  journal = {Monthly Notices of the Royal Astronomical Society},
  year    = {2015},
  volume  = {454},
  number  = {4},
  pages   = {3697--3709},
  doi     = {10.1093/mnras/stv2249}
}

@article{Woodham2021,
  author  = {Woodham, L. D. and Horbury, T. S. and Matteini, L. and Woolley, T. and Laker, R.
             and Bale, S. D. and Nicolaou, G. and Stawarz, J. E. and Stansby, D. and Hietala, H.
             and Larson, D. E. and Livi, R. and Verniero, J. L. and McManus, M. and Kasper, J. C.
             and Korreck, K. E. and Raouafi, N. and Moncuquet, M. and Pulupa, M. P.},
  title   = {Enhanced proton parallel temperature inside patches of switchbacks in the inner heliosphere},
  journal = {Astronomy \& Astrophysics},
  year    = {2021},
  volume  = {650},
  pages   = {L1},
  doi     = {10.1051/0004-6361/202039415}
}

@article{Yogesh2025,
  author  = {Yogesh and Ofman, Leon and Boardsen, Scott A. and Klein, Kristopher G. and
             Martinovi{\'c}, Mihailo and Sadykov, Viacheslav M. and Verniero, Jaye L. and
             Shankarappa, Niranjana and Jian, Lan K. and Mostafavi, Parisa and Huang, Jia and
             Paulson, K. W.},
  title   = {Evidence of Interaction between Ion-scale Waves and Ion Velocity Distributions in the Solar Wind},
  journal = {The Astrophysical Journal},
  year    = {2025},
  month   = jun,
  volume  = {986},
  number  = {2},
  eid     = {119},
  pages   = {119},
  doi     = {10.3847/1538-4357/add467}
}

@article{Reville2018,
  author  = {R{\'e}ville, Victor and Tenerani, Anna and Velli, Marco},
  title   = {Parametric Decay and the Origin of the Low-frequency Alfv{\'e}nic Spectrum of the Solar Wind},
  journal = {The Astrophysical Journal},
  year    = {2018},
  month   = oct,
  volume  = {866},
  number  = {1},
  eid     = {38},
  pages   = {38},
  doi     = {10.3847/1538-4357/aadb8f},
  eprint  = {1806.05762},
  archivePrefix = {arXiv},
  primaryClass  = {astro-ph.SR}
}

@article{Hahn2022,
  author  = {Hahn, Michael and Fu, Xiangrong and Savin, Daniel Wolf},
  title   = {Evidence for Parametric Decay Instability in the Lower Solar Atmosphere},
  journal = {The Astrophysical Journal},
  year    = {2022},
  volume  = {933},
  number  = {1},
  eid     = {52},
  pages   = {52},
  doi     = {10.3847/1538-4357/ac7147},
  eprint  = {2204.09559},
  archivePrefix = {arXiv},
  primaryClass  = {astro-ph.SR}
}

@article{Tenerani2017,
  author  = {Tenerani, Anna and Velli, Marco and Hellinger, Petr},
  title   = {The Parametric Instability of Alfv{\'e}n Waves: Effects of Temperature Anisotropy},
  journal = {The Astrophysical Journal},
  year    = {2017},
  volume  = {851},
  number  = {2},
  eid     = {99},
  pages   = {99},
  doi     = {10.3847/1538-4357/aa9bef},
  eprint  = {1711.06371},
  archivePrefix = {arXiv},
  primaryClass  = {astro-ph.SR}
}

@ARTICLE{Huang2020,
  author  = {Huang, J. and Kasper, J. C. and Vech, D. and Klein, K. G. and Stevens, M.
             and Martinovi{\'c}, M. M. and Alterman, B. L. and {\v{D}}urovcov{\'a}, T.
             and Paulson, K. and Maruca, B. A. and Qudsi, R. A. and Case, A. W.
             and Korreck, K. E. and Jian, L. K. and Velli, M. and Lavraud, B.
             and Hegedus, A. and Bert, C. M. and Holmes, J. and Bale, S. D.
             and Larson, D. E. and Livi, R. and Whittlesey, P. and Pulupa, M.
             and MacDowall, R. J. and Malaspina, D. M. and Bonnell, J. W.
             and Harvey, P. and Goetz, K. and de Wit, T. D.},
  title   = {Proton Temperature Anisotropy Variations in Inner Heliosphere Estimated with the First Parker Solar Probe Observations},
  journal = {Astrophys. J. Suppl. Ser.},
  year    = {2020},
  volume  = {246},
  number  = {2},
  pages   = {70},
  eid     = {70},
  doi     = {10.3847/1538-4365/ab74e0}
}

@article{Marsch1983,
  author  = {Marsch, E. and M{\"u}hlh{\"a}user, K.-H. and Rosenbauer, H. and Schwenn, R.},
  title   = {On the equation of state of solar wind ions derived from {Helios} measurements},
  journal = {Journal of Geophysical Research},
  year    = {1983},
  volume  = {88},
  number  = {A4},
  pages   = {2982--2992},
  doi     = {10.1029/JA088iA04p02982}
}

@article{Marsch1982,
  author  = {Marsch, E. and M{\"u}hlh{\"a}user, K.-H. and Schwenn, R. and Rosenbauer, H. and Pilipp, W. and Neubauer, F. M.},
  title   = {Solar wind protons: Three-dimensional velocity distributions and derived plasma parameters measured between 0.3 and 1 {AU}},
  journal = {Journal of Geophysical Research},
  year    = {1982},
  volume  = {87},
  number  = {A1},
  pages   = {52--72},
  doi     = {10.1029/JA087iA01p00052}
}

@article{Hellinger2011,
  author  = {Hellinger, Petr and Matteini, Lorenzo and {\v{S}}tver{\'a}k, {\v{S}}t{\v{e}}p{\'a}n and Tr{\'a}vn{\'\i}{\v{c}}ek, Pavel M. and Marsch, Eckart},
  title   = {Heating and cooling of protons in the fast solar wind between 0.3 and 1 AU: Helios revisited},
  journal = {Journal of Geophysical Research: Space Physics},
  year    = {2011},
  volume  = {116},
  pages   = {A09105},
  doi     = {10.1029/2011JA016674}
}

@article{Hellinger2013,
  author  = {Hellinger, Petr and Tr{\'a}vn{\'\i}{\v{c}}ek, Pavel M. and {\v{S}}tver{\'a}k, {\v{S}}t{\v{e}}p{\'a}n and Matteini, Lorenzo and Velli, Marco},
  title   = {Proton thermal energetics in the solar wind: {Helios} reloaded},
  journal = {Journal of Geophysical Research: Space Physics},
  year    = {2013},
  volume  = {118},
  number  = {4},
  pages   = {1351--1365},
  doi     = {10.1002/jgra.50107}
}

@article{Matteini2024,
  author  = {Matteini, Lorenzo and Tenerani, Anna and Landi, Simone and Verdini, Andrea and
             Velli, Marco and Hellinger, Petr and Franci, Luca and Horbury, Timothy S. and
             Papini, Emanuele and Stawarz, Julia},
  title   = {Alfv{\'e}nic fluctuations in the expanding solar wind: Formation and radial evolution of spherical polarization},
  journal = {Physics of Plasmas},
  year    = {2024},
  month   = mar,
  volume  = {31},
  number  = {3},
  pages   = {032901},
  eid     = {032901},
  doi     = {10.1063/5.0177754},
  publisher = {American Institute of Physics}
}

@article{GrappinVelli1996,
  author  = {Grappin, R. and Velli, M.},
  title   = {Waves and streams in the expanding solar wind},
  journal = {Journal of Geophysical Research: Space Physics},
  year    = {1996},
  volume  = {101},
  number  = {A1},
  pages   = {425--444},
  doi     = {10.1029/95JA02147}
}

@article{GrappinVelliMangeney1993,
  author  = {Grappin, Roland and Velli, Marco and Mangeney, Andr{\'e}},
  title   = {Nonlinear wave evolution in the expanding solar wind},
  journal = {Physical Review Letters},
  year    = {1993},
  volume  = {70},
  number  = {14},
  pages   = {2190--2193},
  doi     = {10.1103/PhysRevLett.70.2190}
}

@article{DePontieuEtAl2007,
  author  = {De Pontieu, B. and McIntosh, S. W. and Carlsson, M. and Hansteen, V. H.
             and Tarbell, T. D. and Schrijver, C. J. and Title, A. M. and Shine, R. A.
             and Tsuneta, S. and Katsukawa, Y. and Ichimoto, K. and Suematsu, Y.
             and Shimizu, T. and Nagata, S.},
  title   = {Chromospheric Alfv{\'e}nic Waves Strong Enough to Power the Solar Wind},
  journal = {Science},
  year    = {2007},
  volume  = {318},
  number  = {5856},
  pages   = {1574--1577},
  doi     = {10.1126/science.1151747}
}

@article{McIntoshEtAl2011,
  author  = {McIntosh, Scott W. and De Pontieu, Bart and Carlsson, Mats and
             Hansteen, Viggo H. and Boerner, Paul and Goossens, Marcel},
  title   = {Alfv{\'e}nic waves with sufficient energy to power the quiet solar corona and fast solar wind},
  journal = {Nature},
  year    = {2011},
  volume  = {475},
  number  = {7357},
  pages   = {477--480},
  doi     = {10.1038/nature10235}
}

@article{MatthaeusEtAl1999,
  author  = {Matthaeus, W. H. and Zank, G. P. and Oughton, S. and Mullan, D. J. and Dmitruk, P.},
  title   = {Coronal Heating by Magnetohydrodynamic Turbulence Driven by Reflected Low-Frequency Waves},
  journal = {The Astrophysical Journal Letters},
  year    = {1999},
  volume  = {523},
  number  = {1},
  pages   = {L93--L96},
  doi     = {10.1086/312259}
}

@article{DmitrukEtAl2002,
  author  = {Dmitruk, Pablo and Matthaeus, William H. and Milano, Luis J. and Oughton, Sean and Zank, Gary P. and Mullan, David J.},
  title   = {Coronal Heating Distribution Due to Low-Frequency, Wave-driven Turbulence},
  journal = {The Astrophysical Journal},
  year    = {2002},
  volume  = {575},
  number  = {1},
  pages   = {571--577},
  doi     = {10.1086/341188},
  eprint  = {astro-ph/0204347},
  archivePrefix = {arXiv}
}

@article{VerdiniVelli2007,
  author  = {Verdini, Andrea and Velli, Marco},
  title   = {Alfv{\'e}n Waves and Turbulence in the Solar Atmosphere and Solar Wind},
  journal = {The Astrophysical Journal},
  year    = {2007},
  volume  = {662},
  number  = {1},
  pages   = {669--676},
  doi     = {10.1086/510710},
  eprint  = {astro-ph/0702205},
  archivePrefix = {arXiv}
}

@article{PerezChandran2013,
  author  = {Perez, Jean Carlos and Chandran, Benjamin D. G.},
  title   = {Direct Numerical Simulations of Reflection-Driven, Reduced Magnetohydrodynamic Turbulence from the Sun to the Alfv{\'e}n Critical Point},
  journal = {The Astrophysical Journal},
  year    = {2013},
  volume  = {776},
  number  = {2},
  pages   = {124},
  doi     = {10.1088/0004-637X/776/2/124},
  eprint  = {1308.4046},
  archivePrefix = {arXiv}
}

@article{vanBallegooijenAsgariTarghi2016,
  author  = {van Ballegooijen, A. A. and Asgari-Targhi, M.},
  title   = {Heating and Acceleration of the Fast Solar Wind by Alfv{\'e}n Wave Turbulence},
  journal = {The Astrophysical Journal},
  year    = {2016},
  volume  = {821},
  number  = {2},
  pages   = {106},
  doi     = {10.3847/0004-637X/821/2/106},
  eprint  = {1602.06883},
  archivePrefix = {arXiv}
}

@article{vanBallegooijenAsgariTarghi2017,
  author  = {van Ballegooijen, A. A. and Asgari-Targhi, M.},
  title   = {Direct and Inverse Cascades in the Acceleration Region of the Fast Solar Wind},
  journal = {The Astrophysical Journal},
  year    = {2017},
  volume  = {835},
  number  = {1},
  eid     = {10},
  pages   = {10},
  doi     = {10.3847/1538-4357/835/1/10},
  eprint  = {1612.02501},
  archivePrefix = {arXiv},
  primaryClass  = {astro-ph.SR}
}

@article{CranmerVanBallegooijenEdgar2007,
  author  = {Cranmer, Steven R. and van Ballegooijen, Adriaan A. and Edgar, Richard J.},
  title   = {Self-consistent Coronal Heating and Solar Wind Acceleration from Anisotropic Magnetohydrodynamic Turbulence},
  journal = {The Astrophysical Journal Supplement Series},
  year    = {2007},
  volume  = {171},
  number  = {2},
  pages   = {520--551},
  doi     = {10.1086/518001},
  eprint  = {astro-ph/0703333},
  archivePrefix = {arXiv}
}

@article{MeyrandSquireMalletChandran2025,
  author  = {Meyrand, R. and Squire, J. and Mallet, A. and Chandran, B. D. G.},
  title   = {Reflection-driven turbulence in the super-Alfv{\'e}nic solar wind},
  journal = {Journal of Plasma Physics},
  year    = {2025},
  volume  = {91},
  number  = {1},
  pages   = {E29},
  doi     = {10.1017/S0022377824001181},
  eprint  = {2308.10389},
  archivePrefix = {arXiv},
  primaryClass  = {physics.plasm-ph}
}

@article{ChandranEtAl2025b,
  author  = {Chandran, Benjamin D. G. and Sioulas, N. and Bale, S. and Bowen, T. and David, V. and Meyrand, R. and Yerger, E.},
  title   = {Intermittent, reflection-driven, strong imbalanced {MHD} turbulence},
  journal = {Journal of Plasma Physics},
  year    = {2025},
  volume  = {91},
  number  = {2},
  pages   = {E57},
  doi     = {10.1017/S0022377825000194},
  eprint  = {2502.04585},
  archivePrefix = {arXiv},
  primaryClass  = {physics.plasm-ph}
}

@article{vanderHolstEtAl2014,
  author  = {van der Holst, Bart and Sokolov, Igor V. and Meng, Xing and Jin, Meng
             and Manchester, Ward B., IV and T{\'o}th, G{\'a}bor and Gombosi, Tamas I.},
  title   = {Alfv{\'e}n Wave Solar Model ({AWSoM}): Coronal Heating},
  journal = {The Astrophysical Journal},
  year    = {2014},
  volume  = {782},
  number  = {2},
  eid     = {81},
  pages   = {81},
  doi     = {10.1088/0004-637X/782/2/81},
  eprint  = {1311.4093},
  archivePrefix = {arXiv}
}

@article{HeinemannOlbert1980,
  author  = {Heinemann, M. and Olbert, S.},
  title   = {Non-WKB Alfv{\'e}n Waves in the Solar Wind},
  journal = {Journal of Geophysical Research: Space Physics},
  year    = {1980},
  volume  = {85},
  number  = {A3},
  pages   = {1311--1327},
  doi     = {10.1029/JA085iA03p01311}
}

@article{Velli1993,
  author  = {Velli, M.},
  title   = {On the propagation of ideal, linear Alfv{\'e}n waves in radially stratified stellar atmospheres and winds},
  journal = {Astronomy and Astrophysics},
  year    = {1993},
  volume  = {270},
  pages   = {304--314},
  url     = {https://ui.adsabs.harvard.edu/abs/1993A%26A...270..304V/abstract}
}

@article{HowesNielson2013,
  author  = {Howes, G. G. and Nielson, K. D.},
  title   = {Alfv{\'e}n wave collisions, the fundamental building block of plasma turbulence. I. Asymptotic solution},
  journal = {Physics of Plasmas},
  year    = {2013},
  volume  = {20},
  number  = {7},
  pages   = {072302},
  doi     = {10.1063/1.4812805},
  eprint  = {1306.1455},
  archivePrefix = {arXiv}
}

@article{AsgariTarghiEtAl2021,
  author  = {Asgari-Targhi, M. and Asgari-Targhi, A. and Hahn, M. and Savin, D. W.},
  title   = {Effects of Density Fluctuations on Alfv{\'e}n Wave Turbulence in a Coronal Hole},
  journal = {The Astrophysical Journal},
  year    = {2021},
  volume  = {911},
  number  = {1},
  eid     = {63},
  pages   = {63},
  doi     = {10.3847/1538-4357/abe9b4},
  url     = {https://ui.adsabs.harvard.edu/abs/2021ApJ...911...63A/abstract}
}

@article{UsmanovEtAl2018,
  author  = {Usmanov, Arcadi V. and Matthaeus, William H. and Goldstein, Melvyn L. and Chhiber, Rohit},
  title   = {The Steady Global Corona and Solar Wind: A Three-dimensional {MHD} Simulation with Turbulence Transport and Heating},
  journal = {The Astrophysical Journal},
  year    = {2018},
  volume  = {865},
  number  = {1},
  eid     = {25},
  pages   = {25},
  doi     = {10.3847/1538-4357/aad687}
}

@article{ShodaYokoyama2018,
  author  = {Shoda, Munehito and Yokoyama, Takaaki},
  title   = {Anisotropic Magnetohydrodynamic Turbulence Driven by Parametric Decay Instability: The Onset of Phase Mixing and Alfv\'en Wave Turbulence},
  journal = {The Astrophysical Journal Letters},
  year    = {2018},
  volume  = {859},
  number  = {2},
  pages   = {L17},
  doi     = {10.3847/2041-8213/aac50c}
}

@article{ShodaYokoyamaSuzuki2018a,
  author  = {Shoda, Munehito and Yokoyama, Takaaki and Suzuki, Takeru K.},
  title   = {A self-consistent model of the coronal heating and solar wind acceleration including compressible and incompressible heating processes},
  journal = {The Astrophysical Journal},
  year    = {2018},
  volume  = {853},
  pages   = {190},
  doi     = {10.3847/1538-4357/aaa3e1}
}

@article{ShodaEtAl2019,
  author  = {Shoda, Munehito and Suzuki, Takeru Ken and Asgari-Targhi, Mahboubeh and Yokoyama, Takaaki},
  title   = {Three-dimensional Simulation of the Fast Solar Wind Driven by Compressible Magnetohydrodynamic Turbulence},
  journal = {The Astrophysical Journal Letters},
  year    = {2019},
  volume  = {880},
  number  = {1},
  pages   = {L2},
  doi     = {10.3847/2041-8213/ab2b45}
}

@article{ShodaYokoyamaSuzuki2018b,
  author  = {Shoda, Munehito and Yokoyama, Takaaki and Suzuki, Takeru K.},
  title   = {Frequency-dependent {Alfv{\'e}n}-wave Propagation in the Solar Wind: Onset and Suppression of Parametric Decay Instability},
  journal = {The Astrophysical Journal},
  year    = {2018},
  volume  = {860},
  number  = {1},
  pages   = {17},
  doi     = {10.3847/1538-4357/aac218},
  eprint  = {1803.02606},
  archivePrefix = {arXiv},
  primaryClass  = {astro-ph.SR}
}

@article{Matsumoto2021,
  author  = {Matsumoto, Takuma},
  title   = {Full compressible 3D {MHD} simulation of solar wind},
  journal = {Monthly Notices of the Royal Astronomical Society},
  year    = {2021},
  volume  = {500},
  number  = {4},
  pages   = {4779--4787},
  doi     = {10.1093/mnras/staa3533},
  eprint  = {2009.03770},
  archivePrefix = {arXiv},
  primaryClass  = {astro-ph.SR}
}

@book{SagdeevGaleev1969,
  author    = {Sagdeev, R. Z. and Galeev, A. A.},
  title     = {Nonlinear Plasma Theory},
  publisher = {W. A. Benjamin},
  address   = {New York},
  year      = {1969},
  series    = {Frontiers in Physics},
  volume    = {34},
  editor    = {O'Neil, T. M. and Book, D. L.},
  pages     = {122},
  bibcode   = {1969npt..book.....S}
}

@ARTICLE{Derby1978,
  author  = {{Derby}, Jr., N.~F.},
  title   = "{Modulational instability of finite-amplitude, circularly polarized Alfv{\'e}n waves.}",
  journal = {\apj},
  year    = {1978},
  month   = sep,
  volume  = {224},
  pages   = {1013-1016},
  doi     = {10.1086/156451},
  adsurl  = {https://ui.adsabs.harvard.edu/abs/1978ApJ...224.1013D}
}

@ARTICLE{Goldstein1978,
  author  = {{Goldstein}, M.~L.},
  title   = "{An instability of finite amplitude circularly polarized Afv{\'e}n waves.}",
  journal = {\apj},
  year    = {1978},
  month   = jan,
  volume  = {219},
  pages   = {700-704},
  doi     = {10.1086/155829},
  adsurl  = {https://ui.adsabs.harvard.edu/abs/1978ApJ...219..700G}
}

@article{JayantiHollweg1993,
  author  = {Jayanti, Venku and Hollweg, Joseph V.},
  title   = {Parametric instabilities of parallel-propagating Alfv{\'e}n waves: Some analytical results},
  journal = {Journal of Geophysical Research: Space Physics},
  year    = {1993},
  volume  = {98},
  pages   = {19049--},
  doi     = {10.1029/93JA02208}
}

@article{DelZanna2001,
  author  = {Del Zanna, L. and Velli, M. and Londrillo, P.},
  title   = {Parametric decay of circularly polarized Alfv{\'e}n waves: A numerical study},
  journal = {Astronomy \& Astrophysics},
  year    = {2001},
  volume  = {367},
  pages   = {705--718},
  doi     = {10.1051/0004-6361:20000455}
}

@article{HoshinoGoldstein1989,
  author  = {Hoshino, M. and Goldstein, M. L.},
  title   = {Decay instability of a circularly polarized Alfv{\'e}n wave: A numerical simulation},
  journal = {Physics of Fluids B: Plasma Physics},
  year    = {1989},
  volume  = {1},
  number  = {6},
  pages   = {1405--},
  doi     = {10.1063/1.858971}
}

@article{Comisel2019,
  author  = {Comi{\c{s}}el, Horia and Narita, Yasuhito and Motschmann, Uwe},
  title   = {Large-amplitude Alfv{\'e}n waves and their role in heating the solar wind protons},
  journal = {Annales Geophysicae},
  year    = {2019},
  volume  = {37},
  pages   = {835--850},
  doi     = {10.5194/angeo-37-835-2019}
}

@article{TeneraniVelli2013,
  author  = {Tenerani, A. and Velli, M.},
  title   = {Parametric decay of large-amplitude Alfv{\'e}n waves in an expanding solar wind},
  journal = {Journal of Geophysical Research: Space Physics},
  year    = {2013},
  volume  = {118},
  pages   = {7507--},
  doi     = {10.1002/2013JA019293}
}

@article{DelZanna2015,
  author  = {Del Zanna, L. and Matteini, L. and Landi, S. and Verdini, A. and Velli, M.},
  title   = {Parametric decay of parallel and oblique Alfv{\'e}n waves in the expanding solar wind},
  journal = {Journal of Plasma Physics},
  year    = {2015},
  volume  = {81},
  pages   = {325810102},
  doi     = {10.1017/S0022377814000579}
}

@article{MalaraVelli1996,
  author  = {Malara, F. and Velli, M.},
  title   = {Nonlinear evolution of parametric decay of Alfv{\'e}n waves},
  journal = {Physics of Plasmas},
  year    = {1996},
  volume  = {3},
  pages   = {4427--}
}

@article{Hamabata1993,
  author  = {Hamabata, Hiromitsu},
  title   = {Parametric instabilities of finite-amplitude, circularly polarized Alfv{\'e}n waves in an anisotropic plasma},
  journal = {Journal of Plasma Physics},
  year    = {1993},
  volume  = {49},
  number  = {1},
  pages   = {29--39},
  doi     = {10.1017/S0022377800016780}
}

@article{IshizakiIoka2024,
  author  = {Ishizaki, Wataru and Ioka, Kunihito},
  title   = {Parametric decay instability of circularly polarized Alfv{\'e}n waves in magnetically dominated plasmas},
  journal = {Physical Review E},
  year    = {2024},
  volume  = {110},
  pages   = {015205},
  doi     = {10.1103/PhysRevE.110.015205}
}

@article{Chandran2018,
  author  = {Chandran, Benjamin D. G.},
  title   = {Parametric instability, inverse cascade, and the 1/f range of solar-wind turbulence},
  journal = {Journal of Plasma Physics},
  year    = {2018},
  volume  = {84},
  pages   = {905840106},
  doi     = {10.1017/S0022377818000016}
}

@article{Hollweg1994,
  author  = {Hollweg, Joseph V.},
  title   = {Beat, modulational, and decay instabilities of a circularly polarized Alfv{\'e}n wave},
  journal = {Journal of Geophysical Research: Space Physics},
  year    = {1994},
  volume  = {99},
  pages   = {23431--},
  doi     = {10.1029/94JA02185}
}

@article{SuzukiInutsuka2006,
  author  = {Suzuki, Takeru K. and Inutsuka, Shu-ichiro},
  title   = {Solar winds driven by nonlinear low-frequency Alfv{\'e}n waves from the photosphere: Parametric study for fast/slow winds and disappearance of solar winds},
  journal = {Journal of Geophysical Research: Space Physics},
  year    = {2006},
  volume  = {111},
  number  = {A6},
  pages   = {A06101},
  doi     = {10.1029/2005JA011502},
  eprint  = {astro-ph/0511006},
  archivePrefix = {arXiv}
}

@article{GaleevOraevskii1963,
  author  = {Galeev, A. A. and Oraevskii, V. N.},
  title   = {The Stability of Alfv{\'e}n Waves},
  journal = {Soviet Physics Doklady},
  year    = {1963},
  volume  = {7},
  pages   = {988},
  month   = may,
  adsurl  = {https://ui.adsabs.harvard.edu/abs/1963SPhD....7..988G/abstract}
}

@article{WongGoldstein1986,
  author  = {Wong, H. K. and Goldstein, M. L.},
  title   = {Parametric instabilities of circularly polarized {Alfv\'{e}n} waves including dispersion},
  journal = {Journal of Geophysical Research: Space Physics},
  year    = {1986},
  volume  = {91},
  number  = {A5},
  pages   = {5617--5628},
  doi     = {10.1029/JA091iA05p05617}
}

@article{LongtinSonnerup1986,
  author  = {Longtin, M. and Sonnerup, B. U. {\"O}.},
  title   = {Modulation instability of circularly polarized {Alfv\'{e}n} waves},
  journal = {Journal of Geophysical Research: Space Physics},
  year    = {1986},
  volume  = {91},
  number  = {A6},
  pages   = {6816--6824},
  doi     = {10.1029/JA091iA06p06816}
}

@article{VinasGoldstein1991a,
  author  = {Vi{\~n}as, Adolfo F. and Goldstein, Melvyn L.},
  title   = {Parametric instabilities of circularly polarized large-amplitude dispersive {Alfv\'{e}n} waves: excitation of parallel-propagating electromagnetic daughter waves},
  journal = {Journal of Plasma Physics},
  year    = {1991},
  volume  = {46},
  number  = {1},
  pages   = {107--127},
  doi     = {10.1017/S0022377800015981}
}

@article{VinasGoldstein1991b,
  author  = {Vi{\~n}as, Adolfo F. and Goldstein, Melvyn L.},
  title   = {Parametric instabilities of circularly polarized large-amplitude dispersive {Alfv\'{e}n} waves: excitation of obliquely-propagating daughter and side-band waves},
  journal = {Journal of Plasma Physics},
  year    = {1991},
  volume  = {46},
  number  = {1},
  pages   = {129--152},
  doi     = {10.1017/S0022377800015993}
}

@article{Fu2018,
  author  = {Fu, Xiangrong and Li, Hui and Guo, Fan and Li, Xiaocan and Roytershteyn, Vadim},
  title   = {Parametric Decay Instability and Dissipation of Low-frequency Alfv{\'e}n Waves in Low-beta Turbulent Plasmas},
  journal = {The Astrophysical Journal},
  year    = {2018},
  volume  = {855},
  number  = {2},
  pages   = {139},
  doi     = {10.3847/1538-4357/aaacd6},
  eprint  = {1710.04149},
  archivePrefix = {arXiv},
  primaryClass  = {physics.space-ph},
  bibcode = {2018ApJ...855..139F}
}

@article{LiFuDorfman2024,
  author  = {Li, Feiyu and Fu, Xiangrong and Dorfman, Seth},
  title   = {Effects of wave damping and finite perpendicular scale on three-dimensional {Alfv{\'e}n} wave parametric decay in low-beta plasmas},
  journal = {Physics of Plasmas},
  year    = {2024},
  month   = aug,
  volume  = {31},
  number  = {8},
  pages   = {082113},
  doi     = {10.1063/5.0216871},
  eprint  = {2403.08179},
  archivePrefix = {arXiv},
  primaryClass  = {physics.plasm-ph}
}

@article{TerasawaEtAl1986,
  author  = {Terasawa, T. and Hoshino, M. and Sakai, J.-I. and Hada, T.},
  title   = {Decay instability of finite-amplitude circularly polarized {Alfv{\'e}n} waves: A numerical simulation of stimulated {Brillouin} scattering},
  journal = {Journal of Geophysical Research},
  year    = {1986},
  volume  = {91},
  number  = {A4},
  pages   = {4171--4187},
  doi     = {10.1029/JA091iA04p04171},
  bibcode = {1986JGR....91.4171T}
}

@article{NariyukiHada2007,
  author  = {Nariyuki, Yasuhiro and Hada, Tohru},
  title   = {Consequences of finite ion temperature effects on parametric instabilities of circularly polarized {Alfv{\'e}n} waves},
  journal = {Journal of Geophysical Research: Space Physics},
  year    = {2007},
  volume  = {112},
  pages   = {A10107},
  doi     = {10.1029/2007JA012373},
  bibcode = {2007JGRA..11210107N}
}

@article{Araneda1998,
  author  = {Araneda, J. A.},
  title   = {Parametric Instabilities of Parallel-Propagating {Alfv{\'e}n} Waves: Kinetic Effects in the {MHD}-Model},
  journal = {Physica Scripta},
  year    = {1998},
  volume  = {T75},
  pages   = {164--168},
  doi     = {10.1238/Physica.Topical.075a00164},
  bibcode = {1998PhST...75..164A},
  adsurl  = {https://ui.adsabs.harvard.edu/abs/1998PhST...75..164A/abstract}
}

@article{NariyukiHada2006,
  author  = {Nariyuki, Yasuhiro and Hada, Tohru},
  title   = {Kinetically modified parametric instabilities of circularly polarized {Alfv{\'e}n} waves: Ion kinetic effects},
  journal = {Physics of Plasmas},
  year    = {2006},
  volume  = {13},
  number  = {12},
  pages   = {124501},
  doi     = {10.1063/1.2399468},
  eprint  = {physics/0608306},
  archivePrefix = {arXiv},
  primaryClass  = {physics.plasm-ph}
}

@article{UmekiTerasawa1992,
  author  = {Umeki, H. and Terasawa, T.},
  title   = {Decay instability of incoherent {Alfv{\'e}n} waves in the solar wind},
  journal = {Journal of Geophysical Research: Space Physics},
  year    = {1992},
  volume  = {97},
  number  = {A3},
  pages   = {3113--3119},
  month   = mar,
  doi     = {10.1029/91JA02967},
  bibcode = {1992JGR....97.3113U}
}

@inproceedings{MatteiniLandiVelliHellinger2010,
  author    = {Matteini, Lorenzo and Landi, Simone and Velli, Marco and Hellinger, Petr},
  title     = {On the role of wave-particle interactions in the evolution of solar wind ion distribution functions},
  booktitle = {Twelfth International Solar Wind Conference},
  editor    = {Maksimovic, M. and Issautier, K. and Meyer-Vernet, N. and Moncuquet, M. and Pantellini, F.},
  series    = {AIP Conference Proceedings},
  volume    = {1216},
  pages     = {223--226},
  year      = {2010},
  month     = mar,
  publisher = {American Institute of Physics},
  address   = {Melville, NY},
  doi       = {10.1063/1.3396299},
  bibcode   = {2010AIPC.1216..223M}
}

@article{Shi2017,
  author  = {Shi, Mijie and Li, Hui and Xiao, Chijie and Wang, Xiaogang},
  title   = {The Parametric Decay Instability of {Alfv{\'e}n} Waves in Turbulent Plasmas and the Applications in the Solar Wind},
  journal = {The Astrophysical Journal},
  year    = {2017},
  volume  = {842},
  number  = {1},
  pages   = {63},
  doi     = {10.3847/1538-4357/aa71b6},
  eprint  = {1705.03829},
  archivePrefix = {arXiv},
  primaryClass  = {physics.space-ph},
  bibcode = {2017ApJ...842...63S}
}

@article{NariyukiMatsukiyoHada2008,
  author  = {Nariyuki, Yasuhiro and Matsukiyo, Shuichi and Hada, Tohru},
  title   = {Parametric instabilities of large-amplitude parallel propagating {Alfv{\'e}n} waves: 2D {PIC} simulation},
  journal = {New Journal of Physics},
  year    = {2008},
  volume  = {10},
  number  = {8},
  pages   = {083004},
  doi     = {10.1088/1367-2630/10/8/083004}
}

@article{GonzalezInnocentiTenerani2023,
  author  = {Gonz{\'a}lez, C. A. and Innocenti, Maria Elena and Tenerani, Anna},
  title   = {Particle-in-cell simulations of {Alfv{\'e}n} wave parametric decay in a low-beta plasma},
  journal = {Journal of Plasma Physics},
  year    = {2023},
  volume  = {89},
  number  = {2},
  pages   = {905890208},
  month   = apr,
  doi     = {10.1017/S0022377823000120},
  bibcode = {2023JPlPh..89b9008G},
  eprint  = {2301.07646},
  archivePrefix = {arXiv},
  primaryClass  = {physics.plasm-ph}
}

@article{TeneraniEtAl2020,
  author  = {Tenerani, Anna and Velli, Marco and Matteini, Lorenzo and R{\'e}ville, Victor and Shi, Chen and Bale, Stuart D. and Kasper, Justin C. and Bonnell, John W. and Case, Anthony W. and de Wit, Thierry Dudok and Goetz, Keith and Harvey, Peter R. and Klein, Kristopher G. and Korreck, Kelly and Larson, Davin and Livi, Roberto and MacDowall, Robert J. and Malaspina, David M. and Pulupa, Marc and Stevens, Michael and Whittlesey, Phyllis},
  title   = {Magnetic Field Kinks and Folds in the Solar Wind},
  journal = {The Astrophysical Journal Supplement Series},
  year    = {2020},
  month   = feb,
  volume  = {246},
  number  = {2},
  eid     = {32},
  doi     = {10.3847/1538-4365/ab53e1},
  eprint  = {1912.03240},
  archivePrefix = {arXiv},
  primaryClass  = {physics.space-ph},
  bibcode = {2020ApJS..246...32T},
  adsurl  = {https://ui.adsabs.harvard.edu/abs/2020ApJS..246...32T/abstract}
}

@article{MarriottTenerani2024,
  author  = {Marriott, Maile and Tenerani, Anna},
  title   = {Parametric Decay of a Kinked {Alfv{\'e}n} Wave Packet: 3D Magnetohydrodynamic Simulations},
  journal = {The Astrophysical Journal},
  year    = {2024},
  month   = may,
  volume  = {967},
  number  = {1},
  eid     = {19},
  pages   = {19},
  doi     = {10.3847/1538-4357/ad38b9},
  bibcode = {2024ApJ...967...19M},
  adsurl  = {https://ui.adsabs.harvard.edu/abs/2024ApJ...967...19M/abstract}
}

@article{MarriottTenerani2024b,
  author  = {Marriott, Maile and Tenerani, Anna},
  title   = {Parametric Instability of {Alfv{\'e}n} Waves and Wave Packets in Periodic and Open Systems},
  journal = {The Astrophysical Journal},
  year    = {2024},
  month   = nov,
  volume  = {975},
  number  = {2},
  eid     = {232},
  pages   = {232},
  doi     = {10.3847/1538-4357/ad7eb1},
  bibcode = {2024ApJ...975..232M}
}

@article{Komissarov2025,
  author  = {Komissarov, Serguei S.},
  title   = {Parametric instability of {Alfv{\'e}n} wave packets},
  journal = {Monthly Notices of the Royal Astronomical Society},
  year    = {2025},
  volume  = {542},
  number  = {3},
  pages   = {2510--2524},
  doi     = {10.1093/mnras/staf1385},
  eprint  = {2507.10038},
  archivePrefix = {arXiv},
  primaryClass  = {physics.plasm-ph},
  bibcode = {2025MNRAS.542.2510K},
  adsurl  = {https://ui.adsabs.harvard.edu/abs/2025MNRAS.542.2510K/abstract}
}

@article{HasegawaChen1976,
  author  = {Hasegawa, Akira and Chen, Liu},
  title   = {Parametric Decay of ``Kinetic {Alfv{\'e}n} Wave'' and Its Application to Plasma Heating},
  journal = {Physical Review Letters},
  year    = {1976},
  volume  = {36},
  number  = {23},
  pages   = {1362--1365},
  doi     = {10.1103/PhysRevLett.36.1362},
  bibcode = {1976PhRvL..36.1362H},
  adsurl  = {https://ui.adsabs.harvard.edu/abs/1976PhRvL..36.1362H/abstract}
}

@article{Bowen2018,
  author  = {Bowen, Trevor A. and Badman, Samuel and Hellinger, Petr and Bale, Stuart D.},
  title   = {Density Fluctuations in the Solar Wind Driven by {Alfv{\'e}n} Wave Parametric Decay},
  journal = {The Astrophysical Journal Letters},
  year    = {2018},
  month   = feb,
  volume  = {854},
  number  = {2},
  pages   = {L33},
  doi     = {10.3847/2041-8213/aaabbe},
  eprint  = {1712.09336},
  archivePrefix = {arXiv},
  primaryClass  = {physics.space-ph},
  bibcode = {2018ApJ...854L..33B}
}

@article{KasperEtAl2021,
  author  = {Kasper, J. C. and Klein, K. G. and Lichko, E. and Huang, Jia and Chen, C. H. K. and Badman, S. T. and Bonnell, J. and Whittlesey, P. L. and Livi, R. and Larson, D. and Pulupa, M. and Rahmati, A. and Stansby, D. and Korreck, K. E. and Stevens, M. and Case, A. W. and Bale, S. D. and Maksimovic, M. and Moncuquet, M. and Goetz, K. and Halekas, J. S. and Malaspina, D. and Raouafi, Nour E. and Szabo, A. and MacDowall, R. and Velli, Marco and Dudok de Wit, Thierry and Zank, G. P.},
  title   = {Parker Solar Probe Enters the Magnetically Dominated Solar Corona},
  journal = {Physical Review Letters},
  year    = {2021},
  month   = dec,
  volume  = {127},
  number  = {25},
  pages   = {255101},
  doi     = {10.1103/PhysRevLett.127.255101},
  bibcode = {2021PhRvL.127y5101K},
  pmid    = {35029449}
}

@article{HahnEtAl2025,
  author  = {Hahn, Michael and Fu, Xiangrong and Hofmeister, Stefan J. and Huang, Yifan and Koukras, Alexandros and Savin, Daniel Wolf},
  title   = {Velocity and Density Fluctuations in the Quiet Sun Corona},
  journal = {The Astrophysical Journal},
  year    = {2025},
  volume  = {984},
  number  = {1},
  pages   = {69},
  doi     = {10.3847/1538-4357/adc1c0},
  eprint  = {2503.15689},
  archivePrefix = {arXiv},
  primaryClass  = {astro-ph.SR},
  bibcode = {2025ApJ...984...69H}
}

@article{LiDorfmanFu2025,
  author  = {Li, Feiyu and Dorfman, Seth and Fu, Xiangrong},
  title   = {Measuring the growth of {Alfv{\'e}n} wave parametric decay instability using counter-propagating waves: Theory and simulations},
  journal = {Physical Review E},
  year    = {2025},
  volume  = {112},
  number  = {2},
  pages   = {025206},
  month   = aug,
  doi     = {10.1103/4phq-zl19},
  bibcode = {2025PhRvE.112b5206L},
  eprint  = {2507.13590},
  archivePrefix = {arXiv},
  primaryClass  = {physics.plasm-ph}
}

@article{ChewGoldbergerLow1956,
  author  = {Chew, G. F. and Goldberger, M. L. and Low, F. E.},
  title   = {The {Boltzmann} Equation and the One-Fluid Hydromagnetic Equations in the Absence of Particle Collisions},
  journal = {Proceedings of the Royal Society of London. Series A, Mathematical and Physical Sciences},
  year    = {1956},
  volume  = {236},
  number  = {1204},
  pages   = {112--118},
  month   = jul,
  doi     = {10.1098/rspa.1956.0116}
}

@article{Verniero2020,
  title   = {Parker Solar Probe Observations of Proton Beams Simultaneous with Ion-scale Waves},
  author  = {Verniero, J. L. and Larson, D. E. and Livi, R. and Rahmati, A. and McManus, M. D. and Pyakurel, P. Sharma and Klein, K. G. and Bowen, T. A. and Bonnell, J. W. and Alterman, B. L. and Whittlesey, P. L. and Malaspina, David M. and Bale, S. D. and Kasper, J. C. and Case, A. W. and Goetz, K. and Harvey, P. R. and Korreck, K. E. and MacDowall, R. J. and Pulupa, M. and Stevens, M. L. and Dudok de Wit, T.},
  journal = {The Astrophysical Journal Supplement Series},
  volume  = {248},
  number  = {1},
  eid     = {5},
  pages   = {5},
  year    = {2020},
  month   = may,
  doi     = {10.3847/1538-4365/ab86af},
  url     = {https://doi.org/10.3847/1538-4365/ab86af}
}

@article{Huang2023b,
  title        = {The Temperature, Electron, and Pressure Characteristics of Switchbacks: {Parker} Solar Probe Observations},
  author       = {Huang, Jia and Kasper, Justin C. and Larson, Davin E. and McManus, Michael D. and Whittlesey, Phyllis and Livi, Roberto and Rahmati, Ali and Romeo, Orlando M. and Liu, Mingzhe and Jian, Lan K. and Verniero, Jaye L. and Velli, Marco and Badman, Samuel T. and Rivera, Yeimy J. and Niembro, Tatiana and Paulson, Kristoff W. and Stevens, Michael L. and Case, Anthony W. and Bowen, Trevor A. and Pulupa, Marc and Bale, Stuart D. and Halekas, Jasper S.},
  journal      = {The Astrophysical Journal},
  year         = {2023},
  volume       = {954},
  number       = {2},
  eid          = {133},
  pages        = {133},
  doi          = {10.3847/1538-4357/ace694},
  eprint       = {2306.04773},
  archivePrefix= {arXiv},
  primaryClass = {astro-ph.SR}
}

@article{Huang2023a,
  title         = {Parker Solar Probe Observations of High Plasma {$\beta$} Solar Wind from the Streamer Belt},
  author        = {Huang, Jia and Kasper, J. C. and Larson, Davin E. and McManus, Michael D. and Whittlesey, P. and Livi, Roberto and Rahmati, Ali and Romeo, Orlando and Klein, K. G. and Sun, Weijie and van der Holst, Bart and Huang, Zhenguang and Jian, Lan K. and Szabo, Adam and Verniero, J. L. and Chen, C. H. K. and Lavraud, B. and Liu, Mingzhe and Badman, Samuel T. and Niembro, Tatiana and Paulson, Kristoff and Stevens, M. and Case, A. W. and Pulupa, Marc and Bale, Stuart D. and Halekas, J. S.},
  journal       = {The Astrophysical Journal Supplement Series},
  year          = {2023},
  volume        = {265},
  number        = {2},
  eid           = {47},
  pages         = {47},
  doi           = {10.3847/1538-4365/acbcd2},
  eprint        = {2302.07230},
  archivePrefix = {arXiv},
  primaryClass  = {physics.space-ph}
}

@article{Huang2025,
  title        = {The Temperature Anisotropy and Helium Abundance Features of Alfv{\'e}nic Slow Solar Wind Observed by Parker Solar Probe, Helios, and Wind Missions},
  author       = {Huang, Jia and Larson, Davin E. and Ervin, Tamar and Liu, Mingzhe and Ortiz, Oscar and Martinovi{\'c}, Mihailo M. and Huang, Zhenguang and Chasapis, Alexandros and Chu, Xiangning and Alterman, B. L. and Huang, Zesen and Wei, Wenwen and Verniero, J. L. and Jian, Lan K. and Szabo, Adam and Romeo, Orlando and Rahmati, Ali and Livi, Roberto and Whittlesey, Phyllis and Alnussirat, Samer T. and Kasper, Justin C. and Stevens, Michael L. and Bale, Stuart D.},
  journal      = {The Astrophysical Journal Letters},
  year         = {2025},
  volume       = {986},
  number       = {2},
  pages        = {L28},
  doi          = {10.3847/2041-8213/ade0ac},
  eprint       = {2005.12372},
  archivePrefix= {arXiv},
  primaryClass = {physics.space-ph}
}

@article{Short2024,
  title         = {Quiescent Solar Wind Regions in the Near-Sun Environment: Properties and Radial Evolution},
  author        = {Short, Benjamin and Malaspina, David M. and Chasapis, Alexandros and Verniero, Jaye L.},
  journal       = {The Astrophysical Journal},
  year          = {2024},
  volume        = {975},
  number        = {2},
  eid           = {218},
  pages         = {218},
  month         = nov,
  doi           = {10.3847/1538-4357/ad7b13},
  eprint        = {2406.00174},
  archivePrefix = {arXiv}
}

@article{Laker2024,
  title         = {Coherent deflection pattern and associated temperature enhancements in the near-{S}un solar wind},
  author        = {Laker, Ronan and Horbury, T. S. and Woodham, L. D. and Bale, S. D. and Matteini, L.},
  journal       = {Monthly Notices of the Royal Astronomical Society},
  year          = {2024},
  volume        = {527},
  number        = {4},
  pages         = {10440--10447},
  month         = feb,
  doi           = {10.1093/mnras/stad3351},
  eprint        = {2309.13683},
  archivePrefix = {arXiv},
  primaryClass  = {physics.space-ph}
}

@ARTICLE{Zank2022,
  author        = {{Zank}, G.~P. and {Zhao}, L.-L. and {Adhikari}, L. and {Telloni}, D. and {Kasper}, J.~C. and {Stevens}, M. and {Rahmati}, A. and {Bale}, S.~D.},
  title         = "{Turbulence in the Sub-Alfv{\'e}nic Solar Wind}",
  journal       = {\apjl},
  year          = {2022},
  month         = feb,
  volume        = {926},
  number        = {2},
  eid           = {L16},
  pages         = {L16},
  doi           = {10.3847/2041-8213/ac51da},
  archivePrefix = {arXiv},
  eprint        = {2202.02563},
  primaryClass  = {astro-ph.SR},
  adsurl        = {https://ui.adsabs.harvard.edu/abs/2022ApJ...926L..16Z}
}

@article{Zhao2022,
  author  = {Zhao, L.-L. and Zank, G. P. and Telloni, D. and Stevens, M. and Kasper, J. C. and Bale, S. D.},
  title   = {The Turbulent Properties of the Sub-Alfv{\'e}nic Solar Wind Measured by the Parker Solar Probe},
  journal = {The Astrophysical Journal Letters},
  year    = {2022},
  volume  = {928},
  pages   = {L15},
  doi     = {10.3847/2041-8213/ac5fb0},
  url     = {https://doi.org/10.3847/2041-8213/ac5fb0}
}

@article{Bowen2025,
  title   = {Stochastic Heating in the Sub-Alfv{\'e}nic Solar Wind},
  author  = {Bowen, Trevor A. and Ervin, Tamar and Mallet, Alfred and Chandran, Benjamin D. G. and Sioulas, Nikos and Isenberg, Philip A. and Bale, Stuart D. and Squire, Jonathan and Klein, Kristopher G. and Pezzi, Oreste},
  journal = {Physical Review Letters},
  year    = {2025},
  volume  = {135},
  number  = {25},
  pages   = {255201},
  doi     = {10.1103/rxd8-22m9},
  eprint  = {2509.20654},
  archivePrefix = {arXiv},
  primaryClass  = {physics.space-ph}
}

@article{Chandran2013,
  title         = {Stochastic Heating, Differential Flow, and the Alpha-to-proton Temperature Ratio in the Solar Wind},
  author        = {Chandran, B. D. G. and Verscharen, D. and Quataert, E. and Kasper, J. C. and Isenberg, P. A. and Bourouaine, S.},
  journal       = {The Astrophysical Journal},
  year          = {2013},
  volume        = {776},
  number        = {1},
  eid           = {45},
  pages         = {45},
  month         = oct,
  doi           = {10.1088/0004-637X/776/1/45},
  eprint        = {1307.8090},
  archivePrefix = {arXiv},
  primaryClass  = {astro-ph.SR}
}

@article{Meyrand2021,
  title   = {On the violation of the zeroth law of turbulence in space plasmas},
  author  = {Meyrand, R. and Squire, J. and Schekochihin, A. A. and Dorland, W.},
  journal = {Journal of Plasma Physics},
  year    = {2021},
  volume  = {87},
  number  = {3},
  pages   = {535870301},
  month   = jun,
  doi     = {10.1017/S0022377821000489},
  url     = {https://doi.org/10.1017/S0022377821000489}
}

@article{Squire2022,
  title   = {High-frequency heating of the solar wind triggered by low-frequency turbulence},
  author  = {Squire, Jonathan and Meyrand, Romain and Kunz, Matthew W. and Arzamasskiy, Lev and Schekochihin, Alexander A. and Quataert, Eliot},
  journal = {Nature Astronomy},
  year    = {2022},
  volume  = {6},
  number  = {6},
  pages   = {715--723},
  doi     = {10.1038/s41550-022-01624-z}
}

@article{McIntyre2025,
  title     = {Evidence for the helicity barrier from measurements of the turbulence transition range in the solar wind},
  author    = {McIntyre, J. R. and Chen, C. H. K. and Squire, J. and Meyrand, R. and Simon, P. A.},
  journal   = {Physical Review X},
  year      = {2025},
  volume    = {15},
  number    = {3},
  pages     = {031008},
  month     = jul,
  doi       = {10.1103/PhysRevX.15.031008},
  eprint    = {2407.10815},
  archivePrefix = {arXiv},
  primaryClass  = {astro-ph.SR}
}

@article{Panchal2025,
  title   = {Evidence for a Link between Turbulence and the Generation of Ion Cyclotron Waves via the Helicity Barrier Effect in the Solar Wind},
  author  = {Panchal, Utsav and Wicks, Robert T. and Stawarz, Julia E.},
  journal = {The Astrophysical Journal},
  year    = {2025},
  volume  = {983},
  number  = {2},
  eid     = {160},
  pages   = {160},
  month   = apr,
  doi     = {10.3847/1538-4357/adbfff}
}

@article{vanderHolst2022,
  author  = {van der Holst, B. and Huang, J. and Sachdeva, N. and Kasper, J. C. and {Manchester IV}, W. B. and Borovikov, D. and Chandran, B. D. G. and Case, A. W. and Korreck, K. E. and Larson, D. and Livi, R. and Stevens, M. and Whittlesey, P. and Bale, S. D. and Pulupa, M. and Malaspina, D. M. and Bonnell, J. W. and Harvey, P. R. and Goetz, K. and MacDowall, R. J.},
  title   = {Improving the {Alfv\'en} Wave Solar Atmosphere Model Based on Parker Solar Probe Data},
  journal = {The Astrophysical Journal},
  year    = {2022},
  month   = feb,
  volume  = {925},
  number  = {2},
  eid     = {146},
  pages   = {146},
  doi     = {10.3847/1538-4357/ac3d34}
}

@ARTICLE{Chandran2010,
  author        = {{Chandran}, Benjamin D.~G. and {Li}, Bo and {Rogers}, Barrett N. and {Quataert}, Eliot and {Germaschewski}, Kai},
  title         = "{Perpendicular Ion Heating by Low-frequency Alfv{\'e}n-wave Turbulence in the Solar Wind}",
  journal       = {\apj},
  year          = {2010},
  month         = sep,
  volume        = {720},
  number        = {1},
  pages         = {503-515},
  doi           = {10.1088/0004-637X/720/1/503},
  archivePrefix = {arXiv},
  eprint        = {1001.2069},
  primaryClass  = {astro-ph.SR},
  adsurl        = {https://ui.adsabs.harvard.edu/abs/2010ApJ...720..503C}
}

@ARTICLE{Cerri2021,
  author        = {{Cerri}, S.~S. and {Arzamasskiy}, L. and {Kunz}, M.~W.},
  title         = "{On Stochastic Heating and Its Phase-space Signatures in Low-beta Kinetic Turbulence}",
  journal       = {\apj},
  year          = {2021},
  month         = aug,
  volume        = {916},
  number        = {2},
  eid           = {120},
  pages         = {120},
  doi           = {10.3847/1538-4357/abfbde},
  archivePrefix = {arXiv},
  eprint        = {2102.09654},
  primaryClass  = {astro-ph.SR},
  adsurl        = {https://ui.adsabs.harvard.edu/abs/2021ApJ...916..120C}
}

@article{Inhester1990,
  author  = {Inhester, B.},
  title   = {A drift-kinetic treatment of the parametric decay of large-amplitude Alfv{\'e}n waves},
  journal = {Journal of Geophysical Research: Space Physics},
  volume  = {95},
  number  = {A7},
  pages   = {10525--10539},
  year    = {1990},
  month   = jul,
  doi     = {10.1029/JA095iA07p10525}
}

@ARTICLE{Huang2023c,
  author        = {{Huang}, Zesen and {Sioulas}, Nikos and {Shi}, Chen and {Velli}, Marco and {Bowen}, Trevor and {Davis}, Nooshin and {Chandran}, B. D. G. and {Kang}, Ning and {Shi}, Xiaofei and {Huang}, Jia and {Bale}, Stuart D. and {Kasper}, J. C. and {Larson}, Davin E. and {Livi}, Roberto and {Whittlesey}, P. L. and {Rahmati}, Ali and {Paulson}, Kristoff and {Stevens}, M. and {Case}, A. W. and {Dudok de Wit}, Thierry and {Malaspina}, David M. and {Bonnell}, J. W. and {Goetz}, Keith and {Harvey}, Peter R. and {MacDowall}, Robert J.},
  title         = "{New Observations of Solar Wind $1/f$ Turbulence Spectrum from Parker Solar Probe}",
  journal       = {\apjl},
  year          = {2023},
  month         = jun,
  volume        = {950},
  number        = {1},
  eid           = {L8},
  pages         = {L8},
  doi           = {10.3847/2041-8213/acd7f2},
  archivePrefix = {arXiv},
  eprint        = {2303.00843},
  primaryClass  = {astro-ph.SR}
}

@article{Perrone2019,
  title         = {Radial evolution of the solar wind in pure high-speed streams: {HELIOS} revised observations},
  author        = {Perrone, Denise and Stansby, David and Horbury, T. S. and Matteini, Lorenzo},
  journal       = {Monthly Notices of the Royal Astronomical Society},
  year          = {2019},
  volume        = {483},
  number        = {3},
  pages         = {3730--3737},
  month         = mar,
  doi           = {10.1093/mnras/sty3348},
  archivePrefix = {arXiv},
  eprint        = {1810.04014},
  primaryClass  = {physics.space-ph}
}

@ARTICLE{Telloni2021,
  author       = {{Telloni}, Daniele and {Andretta}, Vincenzo and {Antonucci}, Ester and {Bemporad}, Alessandro and {Capuano}, Giuseppe E.
                  and {Fineschi}, Silvano and {Giordano}, Silvio and {Habbal}, Shadia and {Perrone}, Denise and {Pinto}, Rui F. and {Sorriso-Valvo}, Luca
                  and {Spadaro}, Daniele and {Susino}, Roberto and {Woodham}, Lloyd D. and {Zank}, Gary P. and {Romoli}, Marco and {Bale}, Stuart D.
                  and {Kasper}, Justin C. and {Auch{\`e}re}, Fr{\'e}d{\'e}ric and {Bruno}, Roberto and {Capobianco}, Gerardo and {Case}, Anthony W.
                  and {Casini}, Chiara and {Casti}, Marta and {Chioetto}, Paolo and {Corso}, Alain J. and {Da Deppo}, Vania and {De Leo}, Yara
                  and {Dudok de Wit}, Thierry and {Frassati}, Federica and {Frassetto}, Fabio and {Goetz}, Keith and {Guglielmino}, Salvo L.
                  and {Harvey}, Peter R. and {Heinzel}, Petr and {Jerse}, Giovanna and {Korreck}, Kelly E. and {Landini}, Federico and {Larson}, Davin
                  and {Liberatore}, Alessandro and {Livi}, Roberto and {MacDowall}, Robert J. and {Magli}, Enrico and {Malaspina}, David M.
                  and {Massone}, Giuseppe and {Messerotti}, Mauro and {Moses}, John D. and {Naletto}, Giampiero and {Nicolini}, Gianalfredo
                  and {Nistic{\`o}}, Giuseppe and {Panasenco}, Olga and {Pancrazzi}, Maurizio and {Pelizzo}, Maria G. and {Pulupa}, Marc and {Reale}, Fabio
                  and {Romano}, Paolo and {Sasso}, Clementina and {Sch{\"u}hle}, Udo and {Stangalini}, Marco and {Stevens}, Michael L.
                  and {Strachan}, Leonard and {Straus}, Thomas and {Teriaca}, Luca and {Uslenghi}, Michela and {Velli}, Marco and {Verscharen}, Daniel
                  and {Volpicelli}, Cosimo A. and {Whittlesey}, Phyllis and {Zangrilli}, Luca and {Zimbardo}, Gaetano and {Zuppella}, Paola},
  title        = "{Exploring the Solar Wind from Its Source on the Corona into the Inner Heliosphere during the First Solar Orbiter-Parker Solar Probe Quadrature}",
  journal      = {\apjl},
  year         = 2021,
  month        = oct,
  volume       = {920},
  number       = {1},
  eid          = {L14},
  pages        = {L14},
  doi          = {10.3847/2041-8213/ac282f},
  archivePrefix= {arXiv},
  eprint       = {2110.11031},
  primaryClass = {astro-ph.SR},
  url          = {https://ui.adsabs.harvard.edu/abs/2021ApJ...920L..14T}
}

@article{Cranmer2008,
  title        = {Improved Constraints on the Preferential Heating and Acceleration of Oxygen Ions in the Extended Solar Corona},
  author       = {Cranmer, Steven R. and Panasyuk, Alexander V. and Kohl, John L.},
  journal      = {The Astrophysical Journal},
  volume       = {678},
  pages        = {1480--1497},
  year         = {2008},
  doi          = {10.1086/586890},
  eprint       = {0802.0144},
  archivePrefix= {arXiv},
  primaryClass = {astro-ph}
}

@article{Cranmer2020,
  title         = {Updated Measurements of Proton, Electron, and Oxygen Temperatures in the Fast Solar Wind},
  author        = {Cranmer, Steven R.},
  journal       = {Research Notes of the American Astronomical Society},
  year          = {2020},
  month         = dec,
  volume        = {4},
  number        = {12},
  eid           = {249},
  pages         = {249},
  doi           = {10.3847/2515-5172/abd5ae},
  eprint        = {2012.10509},
  archivePrefix = {arXiv},
  primaryClass  = {astro-ph.SR},
  url           = {https://arxiv.org/abs/2012.10509}
}

@article{Telloni2007,
  title   = {Oxygen temperature anisotropy and solar wind heating above coronal holes out to 5 $R_{\odot}$},
  author  = {Telloni, D. and Antonucci, E. and Dodero, M. A.},
  journal = {Astronomy \& Astrophysics},
  year    = {2007},
  volume  = {476},
  number  = {3},
  pages   = {1341--1346},
  doi     = {10.1051/0004-6361:20077660}
}

@article{Whang1973,
  author  = {Whang, Y. C.},
  title   = {Alfv{\'e}n waves in spiral interplanetary field},
  journal = {Journal of Geophysical Research},
  year    = {1973},
  volume  = {78},
  number  = {31},
  pages   = {7221--7228},
  doi     = {10.1029/JA078i031p07221}
}

@article{Maruca2023,
  author       = {Maruca, Bennett A. and Qudsi, Ramiz A. and Alterman, B. L. and Walsh, Brian M. and Korreck, Kelly E. and Verscharen, Daniel and Bandyopadhyay, Riddhi and Chhiber, Rohit and Chasapis, Alexandros and Parashar, Tulasi N. and Matthaeus, William H. and Goldstein, Melvyn L.},
  title        = {The Trans-Heliospheric Survey: Radial trends in plasma parameters across the heliosphere},
  journal      = {Astronomy \& Astrophysics},
  year         = {2023},
  volume       = {675},
  pages        = {A196},
  doi          = {10.1051/0004-6361/202345951},
  url          = {https://doi.org/10.1051/0004-6361/202345951},
  note         = {Published online 24 July 2023}
}
\bibliographystyle{aasjournalv7}

\end{document}